\documentclass[aps,prd,twocolumn,tightenlines,letterpaper,superscriptaddress,nofootinbib,floatfix]{revtex4-2}

\usepackage{amssymb,latexsym}
\usepackage{amsmath,amsbsy,bbm}
\usepackage{epsfig,bm,color,bbold}
\usepackage{graphicx}
\usepackage{xcolor}
\usepackage{rsfso}
\usepackage{natbib}

\usepackage{todonotes}
\usepackage{slashed}

\begin{document}

\def\a{{\alpha}}
\def\be{{\beta}}
\def\d{{\delta}}
\def\D{{\Delta}}
\def\P{{\Pi}}
\def\p{{\pi}}
\def\e{{\varepsilon}}
\def\ep{{\epsilon}}
\def\G{{\Gamma}}
\def\g{{\gamma}}
\def\k{{\kappa}}
\def\l{{\lambda}}
\def\L{{\Lambda}}
\def\m{{\mu}}
\def\n{{\nu}}
\def\o{{\omega}}
\def\O{{\Omega}}
\def\r{{\rho}}
\def\S{{\Sigma}}
\def\s{{\sigma}}
\def\t{{\tau}}
\def\x{{\xi}}
\def\X{{\Xi}}
\def\z{{\zeta}}

\def\sumint{\sum \hskip-1.35em \int_{ \hskip-0.25em {\phantom{a}} } \, \, \,}

\def\ol#1{{\overline{#1}}}
\def\c#1{{\mathcal{#1}}}
\def\b#1{{\bm{#1}}}
\def\eqref#1{{(\ref{#1})}}

\def\wt#1{{\widetilde{#1}}}

\def\ed#1{{\textcolor{magenta}{#1}}}
\def\edd#1{{\textcolor{cyan}{#1}}}

\def\h#1{{\hat{#1}}}
\def\mr#1{{\mathring{#1}}}

%


\author{Prabal~Adhikari}
\email[]{$\texttt{prabal.adhikari.physics@proton.me}$}
 \affiliation{
Physics Department,
        Faculty of Natural Sciences and Mathematics,
        St.~Olaf College,
        Northfield, MN 55057, USA}
\title{Meson Octet in a Uniform Magnetic Field}

\begin{abstract}
Chiral perturbation theory is utilized to construct the renormalized magnetic masses and decay constants of the meson octet at next-to-leading order. While the neutral pion mass decreases identically to two-flavor chiral perturbation theory, the neutral kaon mass remains unaltered by the magnetic field. The renormalized magnetic masses for the charged mesons change identically. Meson decay constants in the axial and vector channels are constructed. Each of the decay constants increase monotonically in the magnetic background. Low-energy theorems -- Gell-Mann-Oakes-Renner relations for the neutral mesons and their generalization for the charged mesons through the pseudoscalar coupling -- are constructed and provide non-trivial crosschecks.
\end{abstract}

\maketitle
\section{Introduction}
\label{sec:introduction}
Magnetic fields are phenomenologically important for strongly interacting systems particularly in astrophysical systems. Surface magnetic fields on magnetars are significant, $10^{10}\,T$, with larger magnetic fields anticipated in the interior of neutron stars~\cite{Harding:2006qn}. Heavy ion collisions at the LHC and the RHIC produce the largest known magnetic fields with fields of size $10^{16}\,T$~\cite{skokov2009estimate,deng2015electric,inghirami2020magnetic,Kharzeev:2007jp,abdulhamid2024observation}. Additionally, large magnetic fields may be important for Big Bang cosmology, in particular the electroweak transition to a confined phase of hadrons~\cite{grasso2001magnetic,vachaspati1991magnetic,vachaspati2021progress,enqvist1993primordial,baym1996magnetic}.

In addition to QCD phenomenology, the impact of strong magnetic fields on the QCD vacuum, phase diagram and hadron properties is of significant theoretical interest~\cite{d2010qcd,d2013susceptibility,bali2014qcd,bali2020magnetic,bali2013magnetic,bruckmann2013inverse,bornyakov2014deconfinement,endrodi2015critical,ding2020qcd,d2021confining,ding2022chiral}. For review articles, see Refs.~\cite{Andersen:2014xxa,shovkovy2013magnetic,kharzeev2013strongly,endrHodi2025qcd,adhikari2025strongly}. Magnetic fields enhance the quark condensate (and the topological susceptibility) as exhibited through model-independent analyses in low-energy QCD~\cite{shushpanov1997quark,agasian2000quark,Cohen:2007bt,Werbos:2007ym,hofmann2020chiral,hofmann2021diamagnetic,hofmann2021thermomagnetic,Adhikari:2021lbl}. Furthermore, they impact the confinement-deconfinement transition, which is understood to be a crossover transition for weak magnetic fields. For asymptotically large magnetic fields, quarks becomes heavy and can be integrated out -- the resulting pure glue theory is anisotropic and possesses a first order deconfinement transition~\cite{cohen2014new}. Magnetic fields also significantly alter the phase diagram of finite density QCD. For finite isospin and baryon chemical potentials, topological solutions are observed including magnetic vortices and chiral soliton lattices~\cite{Adhikari:2015wva,Adhikari:2018fwm,adhikari2022phonon,Brauner:2016pko,Brauner:2021sci,Evans:2022hwr,hamada2026qcd}.

Chiral perturbation theory~\cite{Gasser:1983yg,Gasser:1984gg} allows for the study of the impact of the impact of weak magnetic fields on mesonic matter in a model-independent setting. Since it encodes the long-range behavior of QCD, it can be used to characterize finite volume corrections in a magnetic background~\cite{gasser1988spontaneously} . For weak fields, $\sqrt{eB}\ll 4\p F_{\p}$, finite volume corrections can be significant, particularly for the vacuum magnetization. Additionally spatially inhomogeneous (repeating) structure, for the chiral condensate, is also predicted~\cite{adhikari2023qcd}. Previously, chiral perturbation theory has been used to characterize the alteration of masses and decay constants of both the neutral and charged pions by a background magnetic field~\cite{tiburzi2008hadrons} and thermal bath~\cite{andersen2012chiral,andersen2012thermal}. These are important for the characterization of the weak decay width of both charged and neutral pions~\cite{adhikari2024chiral}. The leptonic decay width for the charged pion undergoes enhancement while the anomalous decay width for the $\p^{0}\rightarrow\g\g$ for the neutral pion is suppressed. In this context, a novel pion decay was proposed in a pioneering lattice study~\cite{bali2018weak} with a model-independent parametrization of the decay constants provided in a model-dependent construction of both charged and neutral pion decay widths~\cite{coppola2020weak,coppola2019pion,coppola2025pi}. Furthermore, properties of the octet baryons including magnetic moments and polarizabilities have been previously characterized~\cite{deshmukh2018octet}. 

In a recent lattice QCD of Ref.~\cite{ding2026chiral}, the zero temperature properties of the pseudoscalar masses and decay constants have been explored at the physical point. The impact of the magnetic field on the Gell-Mann-Oakes-Renner relations for neutral pion and kaons are also  characterized. Three-flavor chiral perturbation theory provides a natural setting to explore the properties of $(2+1)$-flavor QCD. While the properties of the QCD vacuum including the quark condensates and magnetization have been previously explored~\cite{adhikari2023vacuum}, an analysis of magnetic meson masses and decay constants in three-flavor chiral perturbation theory is missing. We consider their impact on the Gell-Mann-Oaker-Renner relations for the meson octet. Additionally, we characterize the pseudoscalar coupling -- refererred to as `the coupling constant of the pseudoscalar density' to the meson octet in Ref.~\cite{Gasser:1983yg} --  as a function of the magnetic background.  A lattice study of hadrons in two-flavor QCD is performed in Ref.~\cite{bali2018meson}. Furthermore, there are several studies of magnetic effects on hadron properties in the Nambu-Jona-Lasinio model~\cite{fayazbakhsh2012properties,abreu2022properties,simonov2016pion,avancini2017pi0,coppola2020magnetic,mei2026spectral} .

The paper is organized as follows: in Sec.~\ref{sec:chiral-lagrangian}, we present the chiral Lagrangian at next-to-leading order necessary for the construction of meson masses, decay constants and the pseudoscalar coupling. In Secs.~\ref{sec:magnetic-mass} and \ref{sec:decay-constant}, we describe the position space construction of the renormalized masses and decay constants.
In Sec.~\ref{sec:low-energy-theorems}, low energy theorems for the octet mesons are discussed -- while the zero-field Gell-Mann-Oakes-Renner relations continue to hold for the neutral mesons, its structure is modified for the charged mesons. 

\section{Chiral Perturbation Theory Ingredients}
\label{sec:chiral-lagrangian}
Construction of the meson magnetic octet mass and decay constants in three-flavor chiral perturbation theory~\cite{Gasser:1984gg} requires accounting for loop effects. We work up to $\c{O}(p^{4})$, which includes the anomalous Wess-Zumino-Witten (WZW) contribution to the chiral Lagrangian~\cite{wess1971consequences,witten1983global}. A scalar-pseudoscalar source, $\chi$, is required to incorporate quark masses and construct the pseudoscalar coupling. A vector source is required to account for the magnetic field while a left source is required to construct the left current. The WZW Lagrangian encodes a vector current that couples the meson to the vacuum only in a background magnetic field.
\subsection{Chiral Lagrangian at $\c{O}(p^{2})$}
The leading $\c{O}(p^{2})$ chiral Lagrangian is a function of the $SU(3)$ meson octet field, $\Sigma$,
\begin{align}
\mathcal{L}_{2}&=\frac{F^{2}}{4}\big\langle\nabla_{\mu}\Sigma\,(\nabla^{\mu}\Sigma)^{\dagger}+\chi\Sigma^{\dagger}+\Sigma\chi^{\dagger}\big\rangle\,.
\end{align}
The angled brackets represent a flavor trace. The scalar-pseudoscalar source, $\chi=2B_{0}(s+ip)$, is required for the inclusion of quark masses and construction of quark condensates and pseudoscalar coupling. The tree-level octet mass follows from the utilization of the coset representation, $\Sigma=\exp\left(\frac{i\phi_{a}\lambda_{a}}{F}\right)$, 
in terms of Gell-Mann matrices, $\lambda_{a}$, and the quark mass matrix, $s=\textrm{diag}(m_{u},m_{d},m_{s})$. The meson octet fields with their corresponding tree-level masses are
\begin{align}
\label{eq:meson-octet-field}
\p^{\pm}&=\tfrac{1}{\sqrt{2}}(\phi_{1}\mp\phi_{2})&
\mr{m}_{\p^{\pm}}^{2}&=B_{0}(m_{u}+m_{d})\\
\p^{0}&=\phi_{3}&
\mr{m}_{\p^{0}}^{2}&=B_{0}(m_{u}+m_{d})\\
K^{\pm}&=\tfrac{1}{\sqrt{2}}(\phi_{4}\mp\phi_{5})&
\mr{m}_{K^{\pm}}^{2}&=B_{0}(m_{u}+m_{s})\\
K^{0},\ol{K}^{0}&=\tfrac{1}{\sqrt{2}}(\phi_{6}\mp\phi_{7})&
\mr{m}_{K^{0}}^{2}&=B_{0}(m_{d}+m_{s})\\
\eta&=\phi_{8}&
\label{eq:eta-field}
\mr{m}_{\eta}^{2}=\tfrac{1}{3}&B_{0}(m_{u}+m_{d}+4m_{s})\,.
\end{align}
We work in the isospin limit. Nevertheless, quark masses, $m_{q_{f}}$, are labeled uniquely, $q_{f}=u,d,s$. At tree-level, the free energy is $\c{F}_0=-B_{0}F^{2}(m_{u}+m_{d}+m_{s})$ with the degenerate quark condensates, $\langle \ol{q}_{f}q_{f}\rangle_{0}=-B_{0}F^{2}$. The tree-level Gell-Mann-Oakes-Renner relation~\cite{gell1968behavior} for the pions is
\begin{align}
\label{eq:GOR-pion}
\mr{m}_{\p^{Q}}^{2}F^{2}&=-\tfrac{1}{2}(m_{u}+m_{d})\langle \bar{u}u+\bar{d}d \rangle_{0}\,.
\end{align}
Analogous Gell-Mann-Oakes-Reiner relations hold for the kaons
\begin{align}
\mr{m}_{K^{\pm}}^{2}F^{2}&=-\tfrac{1}{2}(m_{u}+m_{s})\langle \bar{u}u+\bar{s}s \rangle_{0}\\
\label{eq:GOR-neutral-kaon}
\mr{m}_{K^{0}}^{2}F^{2}&=-\tfrac{1}{2}(m_{d}+m_{s})\langle \bar{d}d+\bar{s}s \rangle_{0}\,.
\end{align}
As we exhibit through explicit calculation of masses and decay constants, in the following sections, only the neutral meson Gell-Mann-Oakes-Renner relations hold in a magnetic background~\cite{agasian2001gell}. The Gell-Mann-Oakes-Renner relation for the charged meson admits a form involving non-magnetic masses at next-to-leading order, as we discuss in Sec.~\ref{sec:low-energy-theorems}.

The covariant derivative in the chiral Lagrangian, required for the inclusion of a vector and axial sources, is
\begin{align}
\nabla_{\mu}\Sigma=\partial_{\mu}\Sigma-iR_{\mu}\Sigma+i\Sigma L_{\mu}\,.
\end{align}
The magnetic field enters through $R_{\mu}=L_{\mu}=-eA_{\mu}\c{Q}$, with $e>0$, where $\c{Q}$ is the quark charge matrix
\begin{align}
\c{Q}=\tfrac{1}{2}\l_{3}+\tfrac{1}{2\sqrt{3}}\l_{8}\,.
\end{align}
The quadratic Lagrangian and left-currents associated with charged meson then depends on the covariant derivative
\begin{align}
\label{eq:covariant-derivative}
D_{\mu}\varphi^{Q}=(\partial_{\mu}+ieQA_{\mu})\varphi^{Q}\,,
\end{align}
with $\varphi=\pi$, $K$ or $\eta$. $Q=\pm1$ and $Q=0$ for the charged and neutral mesons respectively. The superscript for the octet $\eta$ meson field is omitted, as is conventional. The fully antisymmetric gauge for a uniform magnetic field $\vec{B}=B\hat{z}$
\begin{align}
\label{eq:choice-of-gauge}
A_{\mu}=(0,-By,0,0)\,,
\end{align}
with $B>0$, will be utilized in the construction of the charged Green's function in Sec.~\ref{sec:greens-function}.  


\subsection{Chiral Lagrangian at $\c{O}(p^{4})$}
The next-to-leading order Lagrangian is required to renormalize the masses, decay constants and pseudoscalar densities is
\begin{equation}
\begin{split}
\label{eq:L4}
\mathcal{L}_{4}&\supset L_{4}\big\langle\nabla_{\mu}\Sigma(\nabla^{\mu}\Sigma)^{\dagger}\big\rangle\big\langle\chi\Sigma^{\dagger}+\chi^{\dagger}\Sigma\big\rangle\\
&+L_{5}\big\langle\nabla_{\mu}\Sigma(\nabla^{\mu}\Sigma)^{\dagger}(\chi\Sigma^{\dagger}+\chi^{\dagger}\Sigma) \big\rangle\\
&+L_{6}\big\langle\chi\Sigma^{\dagger}+\chi^{\dagger}\Sigma\big\rangle^{2}+L_{7}\big\langle\chi\Sigma^{\dagger}-\chi^{\dagger}\Sigma\big\rangle^{2}\\
&+L_{8}\big\langle \Sigma \chi^{\dagger}\Sigma \chi^{\dagger}+\chi\Sigma^{\dagger}\chi\Sigma^{\dagger} \big\rangle\\
&-iL_{9}\big\langle R_{\mu\nu}\nabla^{\mu}\Sigma(\nabla^{\nu}\Sigma)^{\dagger}+L_{\mu\nu}(\nabla^{\mu}\Sigma)^{\dagger}\nabla^{\nu}\Sigma \big\rangle \\
&+L_{10}\big\langle\Sigma L_{\mu\nu}\Sigma^{\dagger}R^{\mu\nu}\big\rangle\,.
\end{split}
\end{equation}
$L_{6}$ and $L_{7}$ terms are purely zero-field contributions while $L_{4}$ and $L_{5}$ terms are required for wavefunction renormalization -- the magnetic background introduces no further divergences. The $L_{9}$ and $L_{10}$ give rise to purely finite field contributions to the self-energy of charged mesons. $L_{\mu\nu}$ and $R_{\mu\nu}$ are the field strength tensors for left and right sources
\begin{align}
L_{\mu\nu}&=\partial_{\mu}L_{\nu}-\partial_{\nu}L_{\mu}-i[L_{\mu},L_{\nu}]\\
R_{\mu\nu}&=\partial_{\mu}R_{\nu}-\partial_{\nu}R_{\mu}-i[R_{\mu},R_{\nu}]\,.
\end{align}
In a magnetic background, $L_{\mu\nu}=R_{\mu\nu}=-e\,\c{Q}F_{\mu\nu}$.
For renormalization~\cite{Gasser:1984gg}, we utilize dimensional regularization with $d=4-2\epsilon$. In terms of the $\ol{\rm MS}$ scale, $\L$, the divergence in the low-energy constants, $L_{i}$, is
\begin{align}
\label{eq:lec}
L_{i}&=L_{i}^{r}+\Gamma_{i}\lambda&
\lambda&=-\frac{\Lambda^{-2\epsilon}}{2(4\pi)^{2}}\left(\frac{1}{\epsilon}+1\right)\ ,
\end{align}
Utilizing the scale independence of the bare low-energy-constants, the running of the renormalized couplings, $L_{i}^{r}$ is encoded in the relation
\begin{align}
\Lambda\frac{dL_{i}^{r}}{d\Lambda}&=-\frac{\Gamma_{i}}{(4\pi)^{2}}\,.
\end{align}
It is conventional for the values of the renormalized low-energy-constants, $L_{i}^{r}$, to be stated at the scale of the $\rho$ meson mass, $m_{\rho}$. The constants $\Gamma_{i}$ are
\begin{align}
\Gamma_{4}&=\tfrac{1}{8}&
\Gamma_{5}&=\tfrac{3}{8}&
\Gamma_{6}&=\tfrac{11}{144}&
\Gamma_{7}&=0\\
\Gamma_{8}&=\tfrac{5}{48}&
\Gamma_{9}&=\tfrac{1}{4}&
\Gamma_{10}&=-\tfrac{1}{4}\,.
\end{align}

Construction of the vector pion decay constant requires accounting for the chiral anomaly through the WZW Lagrangian,
\begin{equation}
\begin{split}
\label{eq:LWZW}
\c{L}_{\rm WZW}&=\frac{i}{16\p^{2}}\e^{\m\n\r\s}\langle\c{Z}_{\m\n\r\s}\rangle\\
\c{Z}_{\m\n\r\s}&\supset i\partial_{\m}L_{\n}(L_{\r}\S^{\dagger}R_{\s}\S-\S^{\dagger}R_{\r}\S L_{\s})\\
&+i\partial_{\m}R_{\n}(\S L_{\r}\S^{\dagger}R_{\s}-R_{\r}\S L_{\s}\S^{\dagger})\\
&+(\S^{\dagger}\partial_{\mu}\S)(L_{\n}\partial_{\rho}L_{\s}+\partial_{\n}L_{\r}L_{\s})\\
&+\tfrac{1}{2}(\S^{\dagger}\partial_{\mu}\S)(\S^{\dagger}\partial_{\n}R_{\rho}\S L_{\s}+\S^{\dagger} R_{\n}\S\partial_{\r}L_{\s})\\
&-\tfrac{1}{2}(\S\partial_{\m}\S^{\dagger})(\S \partial_{\n} L_{\r} \S^{\dagger}R_{\s}+\S L_{\n}\S^{\dagger}\partial_{\r}R_{\s})\,.
\end{split}
\end{equation}
Only contributions to $\c{Z}_{\m\n\r\s}$ required for the construction of the left current is retained. The convention, $\e^{0123}=1$, is adopted for the epsilon tensor.

\section{Meson Octet Magnetic Mass}
\label{sec:magnetic-mass}
\subsection{Green's Function, $G_{Q}(x',x)$}
\label{sec:greens-function}
The construction of the meson octet magnetic mass requires the \textit{Euclidean} Green's function, $G_{Q}(x',x)$. For neutral mesons ($Q=0$), it is translationally invariant. However, the charged Green's function ($Q=\pm$), as is evident from the defining relation,
\begin{align}
(-D'_{\mu}D'_{\mu}+\mr{m}_{\varphi^{Q}}^{2})G_{Q}(x',x)=\delta^{(4)}(x'-x)\,.
\end{align}
is not. Since the the $\d$-function and the neutral Green's operator is translationally invariant, the neutral Green's function is translationally invariant, $G_{0}(x',x)=G_{0}(x'-x)$. However, the covariant derivative due to the gauge potential, Eq.~\eqref{eq:choice-of-gauge}, is not translationally invariant. Therefore, the charged Green's function, $G_{\pm}(x',x)$ must also be translationally variant to ensure translational invariant of the product. Nevertheless, the charged Green's function is separable into a translationally invariant and variant contributions
\begin{equation}
\begin{split}
G_{Q}(x',x)&=e^{ieB\Delta x_{1}\ol{x}_{2}}\int_{0}^{\infty}\frac{ds}{(4\pi)^{2}s^{2}}\frac{eBs}{\sinh eBs}e^{-\mr{m}_{\varphi}^{2}s}\\
&\times\exp\left[-\tfrac{eB\left(\Delta x_{1}^{2}+\Delta x_{2}^{2}\right)}{4\tanh eBs}-\tfrac{\Delta x_{0}^{2}+\Delta x_{3}^{2}}{4s}\right]\ ,
\end{split}
\end{equation}
with $\Delta x_{\mu}=x'_{\mu}-x_{\mu}$ and $\ol{x}_{\mu}=\tfrac{1}{2}(x'_\mu+x_{\mu})$. The Schwinger integral is notably translationally invariant with all translational variance encoded in the Schwinger phase.

Tadpole contributions to the renormalized mass require the coincident limit of the Green's function,
\begin{align}
\label{eq:D}
G_{Q}(x,x)\equiv \mathcal{D}_{Q}(\mr{m}_{\varphi}^{2})=\mathcal{D}_{0}(\mr{m}_{\varphi}^{2})+\mathcal{D}_{B}(\mr{m}_{\varphi}^{2})\,.
\end{align}
Despite the absence of translational invariance of the charged Green's function, it is coordinate independent. The ultraviolet divergence in the coincident Green's function is handled using dimensional regularization in $d=4-2\e$ dimensions. Using the $\ol{\rm MS}$ scheme with scale $\Lambda$, the $B$-independent contribution to the Green's function is
\begin{align}
\mathcal{D}_{0}(m^{2})&=-\frac{m^{2}}{(4\pi)^{2}}\left[\frac{1}{\epsilon}+1+\log\frac{\Lambda^{2}}{m^{2}}\right]\ .
\end{align}
The magnetic field dependent contribution to the tadpole
\begin{align}
\mathcal{D}_{B}(m^{2})=\frac{eB}{(4\pi)^{2}}\c{I}(\tfrac{m^{2}}{eB})\,,
\end{align}
is a function of the dimensionless Schwinger integral $\c{I}(z)$. It is finite and admits the closed form expression
\begin{align}
\label{eq:IB}
\c{I}(z)=2\log\Gamma(\tfrac{1+z}{2})+z(1-\log\tfrac{z}{2})-\log(2\pi)\,,
\end{align}
in terms of the $\log\Gamma$ function. In the zero field limit, the function vanishes, as required.
\subsection{Two-point Function \& Meson Mass}
The meson octet magnetic mass is extracted from the renormalized two-point function, $\c{G}_{Q}(x',x)$, which, in Schwinger operator language, is defined as
\begin{align}
\c{G}_{Q}(x',x)&=\langle x'|\,\c{G}_{Q}\,|x\rangle\,.
\end{align}
In terms of the bare Green's function, $G_{Q}(x',x)$, the renormalized Green's function is
\begin{align}
\hspace{-12pt}
\langle x'|\,\c{G}_{Q}\,|x\rangle&=\langle x'|\,G_{Q}\,|x\rangle+\smallint_{y}\langle x'|\,G_{Q}\,|y\rangle\c{O}(y)\langle y|G_{Q}|x\rangle\,,
\end{align}
where $\c{O}(y)|y\rangle\langle y|$ is the perturbative correction due to tadpole diagrams
and the integral $\smallint_{y}$ is compact notation for a four-dimensional Euclidean spacetime integral. In operator representation, $\c{G}_{Q}=G_{Q}+G_{Q}\,\c{O}G_{Q}$. However, because $\c{O}$ can be treated perturbatively, a useful identity for the inverse renormalized Green's function follows
\begin{align}
\c{G}_{Q}^{-1}&=G_{Q}^{-1}-\c{O}\,.
\end{align}
In order to extract the wavefunction renormalization factor and the renormalized pion masses in position space, we separate the renormalized Green's functions into terms that contribute to wavefunction renormalization and terms that do not
\begin{equation}
\begin{split}
\c{G}_{Q}(x',x)&=G_{Q}(x',x)+\smallint_{y} G_{Q}(x,y)\,\c{O}_{2}(y)\,G_{Q}(y,x')\\
&+\smallint_{y} D_{\m}G_{Q}(x,y)\,\c{O}_{1}(y)\,D_{\m}G_{Q}(y,x')\,.
\end{split}
\end{equation}
The covariant derivatives act at spacetime location $y$ and reduces to a regular derivative for $Q=0$. $\c{O}_{1}$ and $\c{O}_{2}$ are constructed through Wick contractions involving the interaction Lagrangian -- because the contributions arise through tadpole diagrams (coincident Green's functions), they are spacetime independent. Symmetrization of the first integral via a symmetric integration by parts is necessary to proceed
\begin{equation}
\begin{split}
\c{G}_{Q}(x',x)&=G_{Q}(x',x)+\smallint_{y} G_{Q}(x,y)\,\c{O}_{2}\,G_{Q}(y,x')\\
&-\tfrac{1}{2}\smallint_{y} D_{\m}D_{\m}G_{Q}(x,y)\,\c{O}_{1}\,G_{Q}(y,x')\\
&-\tfrac{1}{2}\smallint_{y} G_{Q}(x,y)\,\c{O}_{1}(y)\,D_{\m}D_{\m}G_{Q}(y,x')\ .
\end{split}
\end{equation}
Its inversion leads to, in operator form,
\begin{align}
\c{G}_{Q}^{-1}=&\,G_{Q}^{-1}+\tfrac{1}{2}\left[\,\overleftarrow{D^{2}}\,\c{O}_{1}+\c{O}_{1}\,\overrightarrow{D^{2}}\,\right]-\c{O}_{2}\\
=&\,(1-\tfrac{1}{2}\c{O}_{1})\,G_{Q}^{-1}\,(1-\tfrac{1}{2}\c{O}_{1})+\mr{m}^{2}_{\varphi}\,\c{O}_{1}-\c{O}_{2}\,,
\end{align}
where $D^{2}=D_{\mu}D_{\mu}$ acts directionally as indicated.
Upon performing a wavefunction renormalization of the meson field,
\begin{align}
\sqrt{Z_{\varphi^{Q}}}=1+\tfrac{1}{2}\c{O}_{1}\,,
\end{align}
the renormalized Green's function and the corresponding renormalized mass follows
\begin{align}
\c{G}_{Q}^{-1}&=\,G_{Q}^{-1}+m_{\varphi^{Q}}^{2}\\
m_{\varphi^{Q}}^{2}&=\mr{m}_{\varphi^{Q}}^{2}(1-\c{O}_{1})-\c{O}_{2}\,.
\end{align}

With the formal structures established, construction of $\c{O}_{1}$ and $\c{O}_{2}$ requires the $\c{O}(p^{2})$ chiral Lagrangian expanded up to quartic order in the meson fields. The quadratic Lagrangian is a function of 
 the masses and covariant derivatives presented in Eqs.~\eqref{eq:meson-octet-field}--\eqref{eq:eta-field} and \eqref{eq:covariant-derivative} respectively. These, in addition to the quartic contribution of the leading order Lagrangian, $\c{L}_{2,4}$ were previously presented in Ref.~\cite{adhikari2023vacuum}\footnote{There is a typo in Eq. (B.1) of Ref.~\cite{adhikari2023vacuum}. While immaterial for the two-loop free energy, it impacts the calculation of renormalized masses. The coefficient in the fourth-to-last line is missing a factor of 2.}. For renormalization, next-to-leading order contribution of $\c{L}_{4}$ containing two meson fields is also required, see $\c{L}_{4,2}$ of Eq. (B.3) in Ref.~\cite{adhikari2023vacuum}.

For charged mesons, contributions from charged tadpoles cancel after accounting for wavefunction renormalization. Therefore, all magnetic field dependence arises through the $L_{9}$ and $L_{10}$ terms.
 The $L_{9}$ term in $\c{L}_{4,2}$ simplifies through integration by parts. Under the assumption of a uniform external field, the expression (in Minkowski space) simplifies to
\begin{align}
-\frac{2L_{9}}{F^{2}}(eF_{\mu\nu})(eF^{\mu\nu})(\pi^{+}\pi^{-}+K^{+}K^{-})\ .
\end{align}
The identity $[D^{\m},D^{\n}]=ieF^{\m\n}$ has been utilized in its construction. The quadratic contributions from the $L_{9}$ and $L_{10}$ terms in the chiral Lagrangian are identical. Since $\G_{9}=-\G_{10}$, the combination $L_{9}+L_{10}$ is finite and scale-independent. It proves efficacious to define the $O(1)$ combination of low-energy-constants
\begin{align}
\label{eq:Lbar}
\ol{L}=4(4\p)^{2}(L^{r}_{9}+L^{r}_{10})\,.
\end{align}
The magnetic masses of the charged mesons are renormalized additively and have an identical structure\footnote{The most precise values of the low-energy-constants, $L^{r}_{9}$ and $L^{r}_{10}$, and corresponding uncertainties are~\cite{bijnens2014mesonic}
\begin{align}
L^{r}_{9}&=6.9(7)\times 10^{-3}&
L^{r}_{10}&=-5.22(6)\times 10^{-3}\,.
\end{align}
Their values are determined through the axial form factor in the decay channel $\p^{+}\rightarrow e^{+}\nu_{e}\gamma$ and the pion electromagnetic radius. The two-flavor counterpart of $\ol{L}$ is $\ol{\ell}$, 
\begin{align}
\ol{\ell}=\tfrac{1}{3}(\ol{l_{6}}-\ol{l_{5}})=1.0(1)\,,
\end{align}
which is known at the ten percent level. This is significantly smaller than the uncertainly in $\ol{L}$. Combining the uncertainties in $L^{r}_{9}$ and $L^{r}_{10}$ in quadrature leads to fractional uncertainty of approximately $40\%$ in $\ol{L}$, more specifically $\ol{L}=1.06(0.44)$ .}
\begin{align}
\label{eq:charged-meson-mass}
m^{2}_{\varphi^{\pm}}(B)&=m_{\varphi^{\pm}}^{2}+\frac{(eB)^{2}}{(4\p F)^{2}}\ol{L}\,.
\end{align}
Additionally, the expression is consistent with the charged mesons not being pseudo-Goldstone bosons -- in the chiral limit and a finite magnetic background, only the first contribution, $m_{\varphi^{\pm}}^{2}$, vanishes~\footnote{In the presence of a magnetic field, the symmetry of the QCD Lagrangian is $U(2)_{L}\times U(2)_{R}$, which is broken down to $U(2)_{V}$ by the ground state. The axial fluctuations around the vacuum are generated by $\l_{3}$, $\tfrac{1}{\sqrt{2}}(\l_{6}\mp\l_{7})$ and $\l_{8}$. The $SU(2)$ subgroup is generated by $\tfrac{1}{\sqrt{2}}(\l_{6}\mp\l_{7})$ and $\tfrac{1}{2}(\l_{3}-\sqrt{3}\l_{8})$ while the $U(1)$ subgroup is generated by $\tfrac{\sqrt{3}}{2}(\sqrt{3}\l_{3}-\l_{8})$.}. The renormalized masses for the charged mesons, however, do not account for Landau levels, labelled by the quantum number $N$ in the magnetic dispersion relation below,
\begin{align}
E_{N,P_{z}}=\sqrt{P_{z}^{2}+m^{2}_{\varphi^{\pm}}(B)+eB(2N+1)}\,.
\end{align}
The spectrum must be interpreted as that of renormalized charged mesons propagating in a classical magnetic background~\cite{adhikari2024chiral}.
For the lowest Landau level, $N=0$, there is a further additive contribution, $eB$, to the mass squared. While both $e$ and $B$ are renormalized, $eB$ is renormalization invariant~\cite{Schwinger:1951nm}.

For the neutral pion, the charged kaon tadpole diagrams cancel exactly. Consequently, only the charged pion tadpole contributes to the renormalized mass
\begin{align}
\label{eq:neutral-pion-mass}
m_{\p^{0}}^{2}(B)=m_{\p^{0}}^{2}\left[1+\frac{eB}{(4\p F)^{2}}\c{I}(\tfrac{\mr{m}_{\p}^{2}}{eB})\right]\,,
\end{align}
with the result identical to that in two-flavor chiral perturbation theory.
For the neutral kaon, on the other hand, all charged tadpole diagrams cancel leading to no magnetic modification at next-to-leading order\footnote{This observation, though unpublished, was first made by Brian Tiburzi. The neutral kaon electric polarizability was studied on the lattice in Ref.~\cite{detmold2009extracting}.}
\begin{align}
\label{eq:neutral-kaon-mass}
m_{K^{0}}^{2}(B)&=m_{K^{0}}^{2}\,.
\end{align}
The next-to-next-leading order corrections to the mass are of relative $\c{O}\left(\frac{(eB)^{2}}{(4\p F)^{4}}\right)$.\footnote{Additionally, there are no finite volume corrections to the neutral kaon magnetic polarizability at next-to-leading order. There are two types of potential sources of finite volume corrections -- the first through the finite volume contribution of each tadpole diagram that contributes in infinite volume, the second through purely finite volume diagrams. Latter arises through interactions of the form
\begin{equation}
\begin{split}
\c{L}_{2,4}&\supset \tfrac{1}{3F^{2}}(D_{\mu}K_{+}\partial^{\mu}K_{0}K_{-}\ol{K}_{0}+D_{\mu}K_{-}\partial^{\mu}\ol{K}_{0}K_{+}K_{0})\\
&-\tfrac{1}{6F^{2}}(D_{\mu}K_{+}\partial^{\mu}\ol{K}_{0}K_{-}K_{0}+D_{\mu}K_{-}\partial^{\mu}K_{0}K_{+}\ol{K}_{0})\,.
\end{split}
\end{equation}
While they do contribute to the two-point function and the zero momentum projected two-point correlation function, upon source averaging contributions of purely finite volume diagrams vanish~\cite{adhikari2023qcd}.
}
The octet eta, however, picks up corrections from both the charged pion and kaon tadpoles
\begin{equation}
\begin{split}
\label{eq:octet-eta-mass}
m_{\eta}^{2}(B)=m_{\eta}^{2}&-\frac{\mr{m}_{\pi}^{2}}{3}\frac{eB}{(4\p F)^{2}}\c{I}(\tfrac{\mr{m}_{\p}^{2}}{eB})\\
&+\frac{2\mr{m}_{K}^{2}}{3}\frac{eB}{(4\p F)^{2}}\c{I}(\tfrac{\mr{m}_{K}^{2}}{eB})\,.
\end{split}
\end{equation}
There are competing contributions: charged pion tadpoles increase the magnetic mass while charged kaon tadpoles reduce the mass. 

\begin{figure}[t]
\centerline{%
\includegraphics[width=.98\columnwidth]{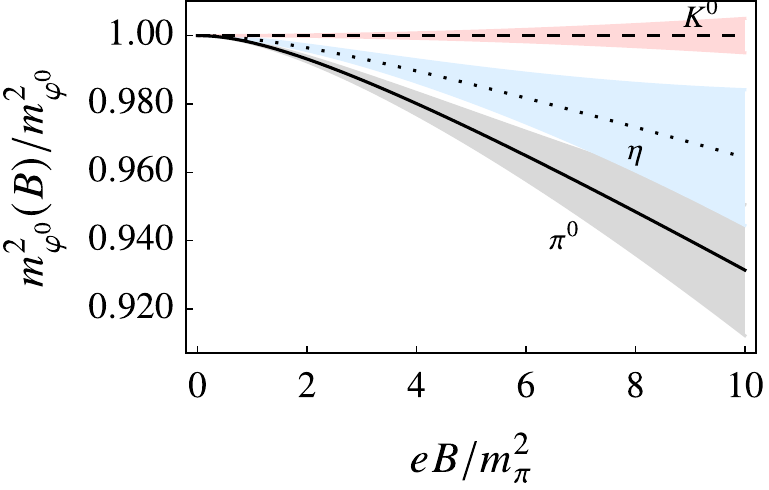}}
\caption{Plot of the magnetic mass of neutral mesons as a function of the magnetic field. The uncertainty bands indicate next-to-next-to-leading order corrections and are generated by varying the relative mass by next-to-next-leading order corrections of size $\pm \frac{(eB)^{2}}{(4\p F)^{4}}$.}  
\label{fig:neutral-meson-mass}
\end{figure}

In Fig.~\ref{fig:neutral-meson-mass}, the neutral meson magnetic masses (normalized by their zero-field values) are plotted as a function of the magnetic field up to $eB=10m_{\p}^{2}$. The neutral pion and octet eta meson magnetic masses both decrease monotonically while the neutral kaon remains massless at next-to-leading order. There is a $7\%$ reduction in the neutral pion mass for the maximum magnetic field plotted while the $\eta$ mass is reduced by approximately $3.5\%$. Since the masses are independent of low-energy-constants, error bands are generated through anticipated next-to-next-to-leading corrections in the chiral expansion. The values of the zero field masses and decay constants   utilized in our analysis~\cite{ParticleDataGroup:2024cfk} are $m_{\p}=138.57\ \texttt{MeV}$, $m_{K}=493.68\ \texttt{MeV}$, $m_{\eta}=547.86\ \texttt{MeV}$, $\sqrt{2}F_{\p}=130.2\ \texttt{MeV}$, and $\sqrt{2}F_{K}=155.7\ \texttt{MeV}$.



\section{Meson Octet Decay Constants}
\label{sec:decay-constant}
The meson decay constant for $\varphi^{Q}$ in a zero magnetic field is defined by the meson-to-vacuum matrix element
\begin{align}
\left.\langle\,0\,|\, J_{L\m}(x)\,|\,\varphi^{Q}(p)\,\rangle\right|_{x=0}=iF_{\varphi^{Q}}\,p_{\m}\ .
\end{align}
The left current, $J_{L\m}(x)$, is defined as the derivative with respect to left-current components, $L^{\mu}=\sum_{a=1}^{8}L^{\mu a}\l^{a}$,
\begin{align}
\label{eq:JL}
J_{L\mu}^{a}=\left.\frac{\partial\c{L}}{\partial L_{}^{\mu a}}\right|_{L_{\mu}=R_{\mu}=-e\c{Q}A_{\mu}}\,.
\end{align}
The right-hand side of the matrix element can be deduced (up to a proportionality factor) by matching the Lorentz structure. Only the four-momentum of the meson field, $p_{\m}$, provides indices with the factor, $F_{\varphi^{Q}}$, identified as the meson decay constant. 

However, in a background magnetic field, four-momentum ceases to be a good quantum number for the charged mesons. It is, therefore, necessary to handle matrix elements in position space for the charged  mesons -- naturally the method is also applicable for the neutral mesons. To this end, consider the correlation function
\begin{align}
\label{eq:correlation-function}
\langle\,0\,|\, J^{\varphi^{Q}}_{L\m}(x')\,\varphi^{Q}(x)\,|\,0\,\rangle\,,
\end{align}
for which Lorentz structure is supplied by the covariant derivative, $D_{\mu}$, Eq.~\eqref{eq:covariant-derivative}. The magnetic field, either through the field strength tensor $F_{\mu\nu}$ or its dual field $\wt{F}_{\mu\nu}=\tfrac{1}{2}\e_{\mu\nu\a\be}F^{\a\be}$ provides further Lorentz indices. 

In addition to Lorentz index matching, parity matching aids the construction of the correlation function. Since the left-current contains both parity odd and even contributions, it is preferable to consider the parity sectors separately. For the axial current, the matrix element is
\begin{equation}
\begin{split}
\label{eq:axial-correlation-function}
\hspace{-10pt}
&\langle\,0\,|\, J^{\varphi^{Q}}_{A\m}(x)\varphi^{Q}(0)\,|\,0\,\rangle=\\
&\ \ \ \ \ \ \ \ \ \ \ -\left[F^{(A1)}_{\varphi^{Q}} D_{\mu}-ieF_{\varphi^{Q}}^{(A2)} F_{\mu\nu}D^{\nu}+\cdots\right]\\
&\ \ \ \ \ \ \ \ \ \ \ \times\langle\,0\,|\,\varphi^{Q}(x)\varphi^{Q}(0)\,|\,0\,\rangle\,.
\end{split}
\end{equation}
$F^{(A1)}_{\varphi^{Q}}$ first appears at tree level while $F^{(A2)}_{\varphi^{\pm}}$ at next-to-leading order.
Since the correlation function is parity odd, the contributions on the right-hand-side must be proportional to the parity odd covariant derivative. Each term may contain one or more of the parity even field strength tensors. More generally, the axial pion decay constant $F^{(An)}_{\varphi^{Q}}$ for $n\ge1$ is associated with the term containing $n-1$ field strength tensors.

The parity even contribution to the correlation function, Eq.~\eqref{eq:correlation-function}, arises through the vector current,
\begin{equation}
\begin{split}
\langle\,0\,|\, &J^{\varphi^{Q}}_{V\m}(x)\,\varphi^{Q}(\,0\,)\,|\,0\,\rangle\\
=&-\left[eF_{\varphi^{Q}}^{(V)}\wt{F}^{\mu\nu}D_{\mu}+\cdots\right]\langle\,0\,|\,\varphi^{Q}(x)\varphi^{Q}(0)\,|\,0\,\rangle
\end{split}
\end{equation}
Since the dual field is parity odd for a magnetic background, the right-hand-side is parity even in spite of the covariant derivative. 
\subsection{Axial Meson Decay Constant, $F^{(A1)}_{\varphi^{Q}}$}
The axial decay constant, $F^{(A1)}_{\varphi^{Q}}$, is most conveniently constructed through the renormalization of the axial current. Accounting for the wavefunction renormalization factor, $Z_{\varphi^{Q}}$, the renormalized axial current admits the structure, $J_{A\mu}{\sim} \left[F\sqrt{Z_{\varphi^{Q}}}+\c{T}_{\varphi^{Q}}+\c{C}_{\varphi^{Q}}\right]D_{\mu}\varphi^{Q}$. 
$\c{T}_{\varphi^{Q}}$ accounts for tadpole contributions while $\c{C}_{\varphi^{Q}}$ accounts for counter-term contributions proportional to $L_{4}$ and $L_{5}$. Since the tadpoles introduce no magnetic field dependent divergences, the resulting magnetic field dependent decay constants are devoid of low-energy-constant contributions. The magnetic corrections are proportional to tadpole contributions of charged mesons 
\begin{align}
\hspace{-5pt}
\frac{F^{(A1)}_{\p^\pm}(B)}{F_{\p}}&=1-\frac{eB}{(4\p F)^{2}}\left[\tfrac{1}{2}\left\{\c{I}(\tfrac{\mr{m}_{\p}^{2}}{eB})+\tfrac{1}{2}\c{I}(\tfrac{\mr{m}_{K}^{2}}{eB})\right\}\right]\\
\label{eq:neutral-pion-FA1}
\hspace{-5pt}
\frac{F^{(A1)}_{\p^0}(B)}{F_{\p}}&=1-\frac{eB}{(4\p F)^{2}}\left[\c{I}(\tfrac{\mr{m}^{2}_{\p}}{eB})+\tfrac{1}{4}\c{I}(\tfrac{\mr{m}^{2}_{K}}{eB})\right]\\
\hspace{-5pt}
\frac{F^{(A1)}_{K^{\pm}}(B)}{F_{K}}&=1-\frac{eB}{(4\p F)^{2}}\left[\tfrac{1}{2}\left\{\tfrac{1}{2}\c{I}(\tfrac{\mr{m}_{\p}^{2}}{eB})+\c{I}(\tfrac{\mr{m}_{K}^{2}}{eB})\right\}\right]\\
\hspace{-5pt}
\label{eq:neutral-kaon-FA1}
\frac{F^{(A1)}_{K^{0}}(B)}{F_{K}}&=1-\frac{eB}{(4\p F)^{2}}\left[\tfrac{1}{4}\left\{\c{I}(\tfrac{\mr{m}_{\p}^{2}}{eB})+\c{I}(\tfrac{\mr{m}_{K}^{2}}{eB})\right\}\right]\\
\hspace{-5pt}
\frac{F^{(A1)}_{\eta}(B)}{F_{\eta}}&=1-\frac{eB}{(4\p F)^{2}}\left[\tfrac{3}{4}\c{I}(\tfrac{\mr{m}_{K}^{2}}{eB})\right]\,.
\end{align}
The magnetic decay constant, $F^{(A1)}_{\varphi^Q}(B)$, has been normalized by the zero field counterparts at next-to-leading order. For each of the pions, the decay constants, $F_{\p}$, are degenerate -- similarly, the kaon decay constants, $F_{K}$. With the exception of the octet $\eta$, for which only charged kaon tadpoles contribute, each magnetic decay constant admits corrections through charged pion and kaon tadpoles.

\begin{figure}[htb]
\centerline{
\includegraphics[width=.98\columnwidth]{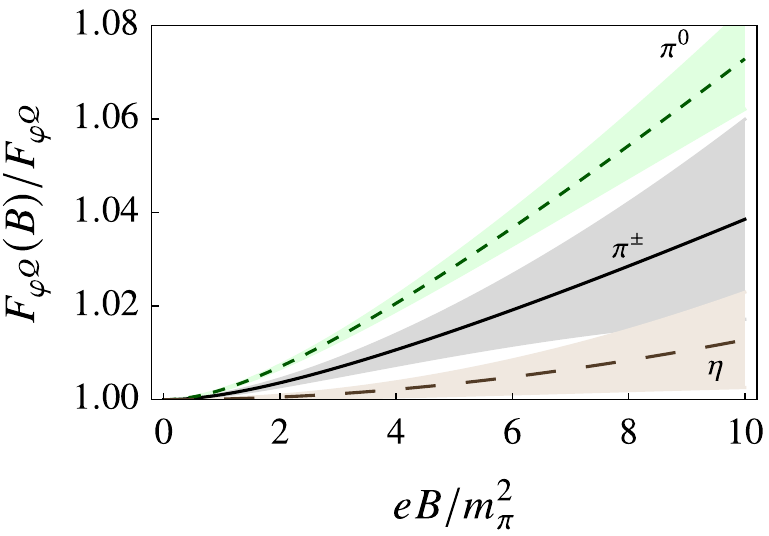}}
\centerline{
\includegraphics[width=.98\columnwidth]{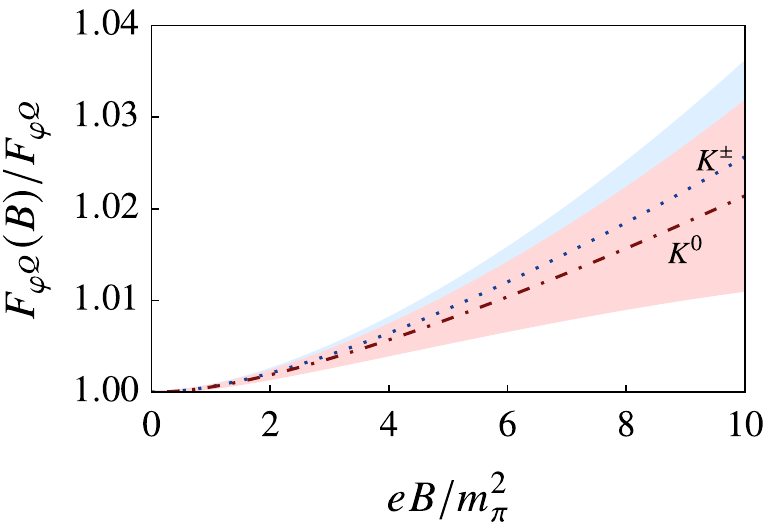}}
\caption{Plots of the axial decay constant, ${F^{(A1)}_{\varphi^Q}(B)}/{F_{\varphi^{Q}}}$ for the meson octet as a function of the magnetic field. (Two plots have been utilized for clarity of presentation.) The uncertainty bands are generated by varying the relative decay constants by next-to-next-leading order corrections of size $\pm \frac{(eB)^{2}}{(4\p F_{\p})^{4}}$. The $K^{\pm}$ decay constant uncertainties are indicated by the light blue shadow while that for $K^{0}$ are indicated by the light red shadow. While the full shadow for the charged kaon is not visible, it is identical in size to the neutral kaon.}  
\label{fig:axial-decay-constant-FA1}
\end{figure}

In Fig.~\ref{fig:axial-decay-constant-FA1}, we plot these ratios as a function of the magnetic field. Each decay constant increases monotonically with the neutral pion exhibiting the largest increase followed by the charged pions, charged kaons, neutral kaons and the octet $\eta$. The increase in the decay constant for the neutral pion is approximately $7\%$ for the largest magnetic field, $eB=10m_{\p}^{2}$, plotted. For the charged pions, the increase is approximately $4\%$ while the increase for the charged kaons is $2.5\%$. For the neutral kaons, the increase is approximately $2.0\%$. The increase for the $\eta$ is modest, approximately $1\%$, due to absence of pion tadpoles.

\subsection{Axial Meson Decay Constant, $F^{(A2)}_{\varphi^{Q}}$}
The magnetic decay constant, $F_{\varphi^{Q}}^{(A2)}$, is non-zero only for the charged mesons. Additionally, it is a pure finite field contribution. As is evident from the correlation function, Eq.~\eqref{eq:correlation-function}, $F_{\varphi^{Q}}^{(A2)}$ arises through a next-to-leading order contribution with all magnetic field dependence entering the correlation function through the explicit $F_{\mu\nu}$. The contribution to the left current arises through the $L_{9}$ and $L_{10}$ terms in the chiral Lagrangian, $\c{L}_{4}$, see Eq.~\eqref{eq:L4},
\begin{equation}
\begin{split}
J_{L\m}^{a}&=-eF_{\m\n}L_{9}\big\langle\l^{a}\left\{(\nabla^{\n}\S)^{\dagger} Q \S+\Sigma^{\dagger} Q \nabla^{\n}\S\right.\\
&\left.-\S^{\dagger}\nabla^{\m}\S Q-Q(\nabla^{\m}\S)^{\dagger}\S\right\}\big\rangle\\
&-2eF_{\m\n}L_{10}\big \langle\l^{a}(\nabla^{\n}\Sigma^{\dagger}Q\Sigma+\Sigma^{\dagger}Q\nabla^{\n}\Sigma)\big \rangle\,.
\end{split}
\end{equation}
The resulting contribution to the left current is non-zero only for the charged mesons. Additionally, their values are identical
\begin{align}
\label{eq:FA2}
\left.J_{L\mu}^{\varphi^{\pm}}\right|_{\rm NLO}&=- F_{\varphi^{\pm}}^{(A2)}F_{\mu\nu}D^{\nu}\varphi^{\pm}&
F_{\varphi^{\pm}}^{(A2)}&=\frac{\ol{L}}{16\p^{2} F}\,.
\end{align}
The combination, $\ol{L}$, defined in Eq.~\eqref{eq:Lbar}, previously appeared in the charged meson magnetic mass, Eq. \eqref{eq:charged-meson-mass}.
\subsection{Vector Meson Decay Constant, $F^{(V)}_{\varphi^{Q}}$}
The contribution to the vector meson decay constant, $F^{(V)}_{\varphi^{Q}}$, arises through the WZW Lagrangian, \eqref{eq:LWZW}. For the charged mesons, the left current at next-to-leading order is gauge covariant 
\begin{align}
\label{eq:JV-WZW}
\left.J_{{L}\m}^{\varphi^{\pm}}\right|_{\rm WZW}&=-eF^{(V)}_{\varphi^{\pm}}\wt{F}_{\m\n}D^{\n}\varphi^{\pm}&
F^{(V)}_{\varphi^{\pm}}&=\frac{1}{8\p^{2}F}\,,
\end{align}
where $\wt{F}_{\m\n}$ is the dual field strength tensor.
For the neutral kaon, the structure is identical with the covariant derivative replaced by a regular one
\begin{align}
\left.J_{L\m}^{K^{0}}\right|_{WZW}&=-eF^{(V)}_{K^{0}}\wt{F}_{\m\n}\partial^{\n}K^{0}&
F^{(V)}_{K^{0}}&=\frac{1}{16\p^{2}F}\,.
\end{align}
Additionally, the vector decay constant is smaller by a factor of two compared to the charged mesons -- the factor is explained at a technical level by the structure of the current, $J_{L}^{a\n}\sim-\frac{3e}{(4\p)^{2}F}\langle\{\l^{a},Q\}\partial^{\n}\phi^{b}\l^{b}\rangle$. Unlike the left current for the neutral kaon, those for the neutral pion and the octet eta, are proportional to the derivative of both the neutral pion and eta fields
\begin{align}
\left.J_{L\m}^{\p^{0}}\right|_{\rm WZW}&=-eF^{(V)}_{\p^{0}}\wt{F}_{\m\n}\left(\partial^{\n}\p^{0}+\sqrt{3}\,\partial^{\n}\eta\right)\\
\left.J_{L\m}^{\eta}\right|_{\rm WZW}&=-eF^{(V)}_{\eta}\wt{F}_{\m\n}\left(\partial^{\n}\eta-\sqrt{3}\,\partial^{\n}\p^{0}\right)\,.
\end{align}
In the absence of isospin breaking, the neutral pion and eta do not mix and therefore the neutral pion-to -vacuum matrix element is independent of the eta contribution to the vector current -- similarly, for the eta-to-vacuum matrix element. The decay constants can then be identified with the respective neutral mesons
\begin{align}
F^{(V)}_{\p^{0}}=F^{(V)}_{\eta}=\frac{1}{8\p^{2}F}\,.
\end{align}
They are identical to that of the charged mesons.
%
%
\section{Low Energy Theorems}
\label{sec:low-energy-theorems}
Low energy theorems provide non-trivial cross-checks for the derived renormalized magnetic masses and decay constants (in the axial channel). Explicit chiral symmetry breaking for the neutral mesons is evident from the divergence of the respective axial current components
\begin{align}
\partial^{\m}J_{A\mu}^{\varphi^{0}}=F^{(A1)}_{\varphi^{0}}(B)m_{\varphi^{0}}^{2}(B)\,\varphi^{0}\,.
\end{align}
The corresponding relationship in QCD for the neutral mesons is
\begin{align}
\label{eq:axial-divergence}
\partial^{\m}\left(\ol{q}\g_{\m}\g_{5}\tfrac{1}{2}\l_{a}q\right)=i\ol{q}\g_{5}\left\{\tfrac{1}{2}\l_{a},s\right\}q\,,
\end{align}
where $s$ is the quark mass matrix and $q(x)$ is the quark field $q^{T}(x)=\big(\,u(x),d(x),s(x)\,\big)$.
Comparison of the correlation function in chiral perturbation theory
\begin{equation}
\begin{split}
\langle\,0\,|\,\partial^{\m}J_{A\mu}^{\varphi^{0}}(x)&\varphi^{0}(0)\,|\,0\,\rangle\\
=&F^{(A1)}_{\varphi^{0}}(B)m_{\varphi^{0}}^{2}(B)\langle\,0\,|\,\varphi^{0}(x)\varphi^{0}(0)\,|\,0\,\rangle
\end{split}
\end{equation}
with the corresponding relation in QCD leads to the low energy theorems
\begin{align}
F_{\p^{0}}(B)m_{\p^{0}}^{2}(B)&=\tfrac{1}{2}(m_{u}+m_{d})G_{\p^{0}}(B)\\
F_{K^{0}}(B)m_{K^{0}}^{2}(B)&=\tfrac{1}{2}(m_{d}+m_{s})G_{K^{0}}(B)\,.
\end{align}
For compactness, we have adopted the notation $F^{(A1)}_{\varphi^{0}}(B)=F_{\varphi^{0}}(B)$.\footnote{An analogous relation for the octet eta
\begin{align}
F^{(A1)}_{\eta}(B)m_{\eta}^{2}(B)&=\tfrac{1}{6}(m_{u}+m_{d}+4m_{s})G_{\eta}(B)\,,
\end{align}
does not appear to hold in chiral perturbation theory. Unlike the magnetic corrections to the masses of the neutral pion and the neutral kaons, the magnetic corrections to the octet eta mass (squared) is not the product of the zero field mass (squared) and a magnetic tadpole. Amusingly, however, in the limit of equal quark masses Eq.~\eqref{eq:octet-eta-mass} reduces to
\begin{equation}
\begin{split}
m_{\eta}^{2}(B)=m_{\eta}^{2}\left[1+\frac{eB}{(4\p F)^{2}}\left\{-\tfrac{1}{3}\c{I}(\tfrac{\mr{m}_{\p}^{2}}{eB})+\tfrac{2}{3}\c{I}(\tfrac{\mr{m}_{K}^{2}}{eB})\right\}\right]\,.
\end{split}
\end{equation}
The coefficients of $\c{I}$ are then precisely the ones required to satisfy the low-energy theorem for the eta. The renormalized pseudoscalar coupling is found to be
\begin{align}
\frac{G_{\eta}(B)}{G_{\eta}}&=1-\frac{eB}{(4 \p F)^{2}}\left[\tfrac{1}{3}\c{I}(\tfrac{\mr{m}_{\p}^{2}}{eB})+\tfrac{1}{12}\c{I}(\tfrac{\mr{m}_{K}^{2}}{eB})\right]\,.
\end{align}
The coefficient in the magnetic eta mass of $-\frac{1}{3}$ for magnetic pion tadpole is consistent with the coefficient above and the coefficient of $\frac{2}{3}$ for the magnetic kaon tadpole multiplies with that in the axial decay constant to give $-\frac{1}{12}$, which is precisely the factor present in the pseudoscalar coupling.
} 
The pseudoscalar coupling, $G_{\varphi^{0}}(B)$, is defined in terms of the pseudoscalar matrix element
\begin{align}
\hspace{-5pt}
\langle\,0\,|\,P_{\varphi^{0}}(x)\varphi^{0}(0)\,|\,0\,\rangle=G_{\varphi^{0}}(B)\langle\,0\,|\,\varphi^{0}(x)\varphi^{0}(0)\,|\,0\,\rangle\,,
\end{align}
with the corresponding pseudoscalar operator in QCD defined as $P_{a}(x)=i\ol{q}(x)\g_{5}\l_{a}q(x)$.  The couplings are readily constructed through the pseudoscalar current~\cite{Gasser:1983yg}
\begin{align}
J_{Pa}=\left.\frac{\partial\c{L}}{\partial p_{a}}\right|_{p=0}\,,
\end{align}
which is proportional to the corresponding pseudoscalar fields, $J_{P}\sim G_{\varphi}\varphi$. The resulting relations
\begin{align}
\frac{G_{\p^{0}}(B)}{G_{\p}}&=1-\frac{eB}{(4\p F)^{2}}\left[\tfrac{1}{4}\c{I}(\tfrac{\mr{m}_{K}^{2}}{eB})\right]\\
\frac{G_{K^{0}}(B)}{G_{K}}&=1-\frac{eB}{(4\p F)^{2}}\left[\tfrac{1}{4}\left\{\c{I}(\tfrac{\mr{m}_{\p}^{2}}{eB})+\c{I}(\tfrac{\mr{m}_{K}^{2}}{eB})\right\}\right]
\end{align}
are consistent as is readily verified using the renormalized meson masses and decay constants. For the neutral pion, the contribution of the charged pion tadpole cancels in the product $F_{\p^{0}}(B)m_{\p^{0}}^{2}(B)$ consistent with the expression for $G_{\p^{0}}(B)$. For the neutral kaon, since the mass is not modified, the modification of the axial decay constant and the pseudoscalar coupling are identical. 

Additionally, the Gell-Mann-Oakes-Renner relations must be obeyed for the neutral pion, Eq.~\eqref{eq:GOR-pion} and the neutral kaon, Eq.~\eqref{eq:GOR-neutral-kaon}. The shift in the quark condensates in the magnetic background is deduced through the magnetic contribution to the free energy, which is due to the charged mesons. The shift in the up-quark condensate arises through the charged pions and kaons, that of the down-quark condensate arises through the charged pion and that of the strange-quark condensate arises through the charged kaons
\begin{align}
\frac{\langle\,\ol{u}u\,\rangle_{B}}{\langle\,\ol{u}u\,\rangle_{0}}&=-\frac{eB}{(4\p F)^{2}}\left[\c{I}(\tfrac{\mr{m}^{2}_{\p}}{eB})+\c{I}(\tfrac{\mr{m}^{2}_{K}}{eB})\right]\\
\frac{\langle\,\ol{d}d\,\rangle_{B}}{\langle\,\ol{d}d\,\rangle_{0}}&=-\frac{eB}{(4\p F)^{2}}\c{I}(\tfrac{\mr{m}^{2}_{\p}}{eB})\\
\frac{\langle\,\ol{s}s\,\rangle_{B}}{\langle\,\ol{s}s\,\rangle_{0}}&=-\frac{eB}{(4\p F)^{2}}\c{I}(\tfrac{\mr{m}^{2}_{K}}{eB})\,.
\end{align}
$\langle\, \ol{q}_{f}q_{f}\,\rangle_{0}=-B_{0}F^{2}$ are the tree-level quark condensates.
A low-energy theorem $\langle\,\ol{d}d\,\rangle_{B}+\langle\,\ol{s}s\,\rangle_{B}=\langle\,\ol{u}u\,\rangle_{B}$, naturally follows~\cite{Adhikari:2021lbl}. It holds up to next-to-leading order and is violated by next-to-next-to-leading order corrections~\cite{adhikari2023vacuum}. Utilizing the neutral pion mass, Eq.~\eqref{eq:neutral-pion-mass} and the axial decay constant, Eq.~\eqref{eq:neutral-pion-FA1}, we find that
\begin{equation}
\begin{split}
&\frac{m_{\p^{0}}^{2}(B)F_{\p^{0}}^{2}(B)}{m_{\p^{0}}^{2}F_{\p^{0}}^{2}}=1+\frac{\langle\, \ol{u}u+\ol{d}d\,\rangle_{B}}{\langle\, \ol{u}u+\ol{d}d\,\rangle_{0}}\,.
\end{split}
\end{equation}
In the limit, $B=0$, the right-hand-side reduces to one. Using the neutral kaon decay constant and the observation that the neutral kaon mass is unaltered by the magnetic field at next-to-leading order, the relations
\begin{align}
&\frac{m_{K^{0}}^{2}(B)F_{K^{0}}^{2}(B)}{m_{K^{0}}^{2}F_{K^{0}}^{2}}=\frac{F_{K^{0}}^{2}(B)}{F_{K^{0}}^{2}}=1+\frac{\langle \,\ol{d}d+\ol{s}s\,\rangle_{B}}{\langle\, \ol{d}d+\ol{s}s\,\rangle_{0}}\,,
\end{align}
exhibit the Gell-Mann-Oakes-Renner relation for the neutral kaon.

For charged mesons, the divergence of the axial current is defined covariantly
\begin{align}
D^{\m}J_{A\m}^{\varphi^{\pm}}&=-\left[F_{\varphi^{\pm}}^{(A1)}D_{\m}D^{\m}+\tfrac{1}{2}e^{2}F_{\varphi^{\pm}}^{(A2)}F_{\m\n}F^{\m\n}\right]\varphi^{\pm}
\end{align}
with the corresponding relation from QCD arising upon the replacement of the derivative with a covariant one, $\partial_{\m}\rightarrow\c{D}_{\m}=\partial_{\m}+ie\c{Q}A_{\m}$, in Eq.~\eqref{eq:axial-divergence}.
The resulting correlation function
\begin{equation}
\begin{split}
\langle\,0\,|\,D^{\m}J_{A\m}^{\varphi^{\pm}}(x)&\varphi^{\pm}(0)\,|\,0\,\rangle\\
=&F_{\varphi^{\pm}}(B)m_{\varphi^{\pm}}^{2}\,\langle\,0\,|\,\varphi^{\pm}(x)\varphi^{\pm}(0)\,|\,0\,\rangle\,,
\end{split}
\end{equation}
follows upon utilizing the Green's function relation. $F_{\varphi^{\pm}}(B)$ is defined through the relation
\begin{align}
F_{\varphi^{\pm}}(B)m_{\varphi^{\pm}}^{2}=F^{(A1)}_{\varphi^{\pm}}(B)\,m^{2}_{\varphi^{\pm}}(B)-F_{\varphi^{\pm}}^{(A2)}(eB)^{2}\,.
\end{align}
The contribution proportional to $F_{\varphi^{\pm}}^{(A2)}$, see Eq.~\eqref{eq:FA2}, cancels with that from the magnetic mass, $m_{\varphi^{\pm}}^{2}(B)$.  Consequently, the low energy theorems for charged mesons are modified
\begin{align}
F_{\p^{\pm}}(B)m_{\p^{\pm}}^{2}&=\tfrac{1}{2}(m_{u}+m_{d})G_{\p^{\pm}}(B)\\
F_{K^{\pm}}(B)m_{K^{\pm}}^{2}&=\tfrac{1}{2}(m_{d}+m_{s})G_{K^{\pm}}(B)\,,
\end{align}
and consistent with the magnetic pseudoscalar densities
\begin{align}
\hspace{-5pt}
\frac{G_{\p^{\pm}}(B)}{G_{\p}}&=1-\frac{eB}{(4\p F)^{2}}\left[\tfrac{1}{2}\left\{\c{I}(\tfrac{\mr{m}_{\p}^{2}}{eB})+\tfrac{1}{2}\c{I}(\tfrac{\mr{m}_{K}^{2}}{eB})\right\}\right]\\
\hspace{-5pt}
\frac{G_{K^{\pm}}(B)}{G_{K}}&=1-\frac{eB}{(4\p F)^{2}}\left[\tfrac{1}{2}\left\{\tfrac{1}{2}\c{I}(\tfrac{\mr{m}_{\p}^{2}}{eB})+\c{I}(\tfrac{\mr{m}_{K}^{2}}{eB})\right\}\right].
\end{align}
Moreover, the Gell-Mann-Oakes-Renner relation for the charged mesons generalizes,
\begin{align}
\frac{m_{\p^{\pm}}^{2}F^{2}_{\p^{\pm}}(B)}{m_{\p}^{2}F_{\p}^{2}}&=\frac{G^{2}_{\p^{\pm}}(B)}{G^{2}_{\p}}=1+\frac{\langle\, \ol{u}u+\ol{d}d\,\rangle_{B}}{\langle\, \ol{u}u+\ol{d}d\,\rangle_{0}}\\
\frac{m_{K^{\pm}}^{2}F^{2}_{K^{\pm}}(B)}{m_{K}^{2}F_{K}^{2}}&=\frac{G^{2}_{K^{\pm}}(B)}{G^{2}_{K}}=1+\frac{\langle \,\ol{u}u+\ol{s}s\,\rangle_{B}}{\langle\, \ol{u}u+\ol{s}s\,\rangle_{0}}\,.
\end{align}
The quark condensate ratio is identical to that of the neutral counterpart for the charged pion. For the charged kaon, however, the condensate ratio involves the up-quark condensate. 
\section{Conclusion}
\label{sec:conclusion}
In this paper, we have characterized the impact of magnetic fields on the meson octet masses and decay constants using three-flavor chiral perturbation theory. There are some surprises -- at next-to-leading order the magnetic mass for the neutral pion is identical to the result in two-flavor chiral perturbation theory and the magnetic field does not modify the mass of the neutral kaon and finite volume corrections in the neutral kaon polarizability is absent at this order. The magnetic mass of the octet eta decreases monotonically with the magnetic field. For the charged mesons, the corrections to the renormalized mass and pole mass are identical. The first axial decay constant, $F^{(A1)}_{\varphi^{Q}}$, monotonically increases for each of the mesons with the neutral pion exhibit the most significant increase, followed by the charged pions, charged kaons, neutral kaons and the octet eta. The second axial decay constant, $F^{(A2)}_{\varphi^{\pm}}$, is non-zero for the charged mesons and degenerate. The vector decay constant, $F^{(V)}_{\varphi^{Q}}$ is identical for each of the mesons with the exception of the neutral kaon for which it is smaller by a factor of two.

For the neutral mesons, Gell-Mann-Oakes relations provide non-trivial tests of the validity of the magnetic masses and decay constants -- their alteration must be consistent with the enhancement of the corresponding quark condensates. Additionally, the relations generalize for the charged mesons with the replacement of the magnetic mass by the renormalized mass at zero field. 

The renormalized magnetic masses and decay constants are inputs in the construction of magnetic decay widths, which may be explored further in future work, particularly for the kaons. For the pions, decay width modifications were constructed in two-flavor chiral perturbation theory in Ref.~\cite{adhikari2024chiral} for the anomalous channel, $\p^{0}\rightarrow\g\g$ and the non-anomalous leptonic channel $\p^{+}\rightarrow \ell^{+}\n_{\ell}$. Among the inputs required for the analysis, only $F^{(A1)}_{\p^{\pm}}(B)$ is modified in three-flavor chiral perturbation theory. The ratio of the three-flavor axial decay constant to its two-flavor counterpart
\begin{align}
\frac{\left.F^{(A1)}_{\p^{\pm}}(B)\right|_{N_{f}=3}}{\left.F^{(A1)}_{\p^{\pm}}(B)\right|_{N_{f}=2}}=-\frac{eB}{(4\p F)^{2}}\tfrac{1}{4}\c{I}(\tfrac{\mr{m}_{K}^{2}}{eB})
\end{align}
suggests a $0.3\%$ larger value of the decay constant at $eB=10m_{\p}^{2}$. Consequently, the charged decay width in three-flavor chiral perturbation theory is not significantly modified by the inclusion of the strange mass. Further directions for exploration include the study of finite temperature effects on the octet masses and decay constants.
\section*{Acknowledgements}
P.A. is supported by the U.S. Department of Energy, Office of Science, Office of Nuclear Physics and Quantum Horizons Program under Award Number 
DE-SC0024385. P.A. acknowledges discussions with Brian Tiburzi.
\bibliographystyle{apsrev4-2}
\bibliography{/Users/prabal121/Desktop/ResearchC/bibly}

\begin{thebibliography}{73}%
\makeatletter
\providecommand \@ifxundefined [1]{%
 \@ifx{#1\undefined}
}%
\providecommand \@ifnum [1]{%
 \ifnum #1\expandafter \@firstoftwo
 \else \expandafter \@secondoftwo
 \fi
}%
\providecommand \@ifx [1]{%
 \ifx #1\expandafter \@firstoftwo
 \else \expandafter \@secondoftwo
 \fi
}%
\providecommand \natexlab [1]{#1}%
\providecommand \enquote  [1]{``#1''}%
\providecommand \bibnamefont  [1]{#1}%
\providecommand \bibfnamefont [1]{#1}%
\providecommand \citenamefont [1]{#1}%
\providecommand \href@noop [0]{\@secondoftwo}%
\providecommand \href [0]{\begingroup \@sanitize@url \@href}%
\providecommand \@href[1]{\@@startlink{#1}\@@href}%
\providecommand \@@href[1]{\endgroup#1\@@endlink}%
\providecommand \@sanitize@url [0]{\catcode `\\12\catcode `\$12\catcode
  `\&12\catcode `\#12\catcode `\^12\catcode `\_12\catcode `\%12\relax}%
\providecommand \@@startlink[1]{}%
\providecommand \@@endlink[0]{}%
\providecommand \url  [0]{\begingroup\@sanitize@url \@url }%
\providecommand \@url [1]{\endgroup\@href {#1}{\urlprefix }}%
\providecommand \urlprefix  [0]{URL }%
\providecommand \Eprint [0]{\href }%
\providecommand \doibase [0]{https://doi.org/}%
\providecommand \selectlanguage [0]{\@gobble}%
\providecommand \bibinfo  [0]{\@secondoftwo}%
\providecommand \bibfield  [0]{\@secondoftwo}%
\providecommand \translation [1]{[#1]}%
\providecommand \BibitemOpen [0]{}%
\providecommand \bibitemStop [0]{}%
\providecommand \bibitemNoStop [0]{.\EOS\space}%
\providecommand \EOS [0]{\spacefactor3000\relax}%
\providecommand \BibitemShut  [1]{\csname bibitem#1\endcsname}%
\let\auto@bib@innerbib\@empty
\bibitem [{\citenamefont {Harding}\ and\ \citenamefont
  {Lai}(2006)}]{Harding:2006qn}%
  \BibitemOpen
  \bibfield  {author} {\bibinfo {author} {\bibfnamefont {A.~K.}\ \bibnamefont
  {Harding}}\ and\ \bibinfo {author} {\bibfnamefont {D.}~\bibnamefont {Lai}},\
  }\href {https://doi.org/10.1088/0034-4885/69/9/R03} {\bibfield  {journal}
  {\bibinfo  {journal} {Rept. Prog. Phys.}\ }\textbf {\bibinfo {volume} {69}},\
  \bibinfo {pages} {2631} (\bibinfo {year} {2006})},\ \Eprint
  {https://arxiv.org/abs/astro-ph/0606674} {arXiv:astro-ph/0606674 [astro-ph]}
  \BibitemShut {NoStop}%
\bibitem [{\citenamefont {Skokov}\ \emph {et~al.}(2009)\citenamefont {Skokov},
  \citenamefont {Illarionov},\ and\ \citenamefont
  {Toneev}}]{skokov2009estimate}%
  \BibitemOpen
  \bibfield  {author} {\bibinfo {author} {\bibfnamefont {V.}~\bibnamefont
  {Skokov}}, \bibinfo {author} {\bibfnamefont {A.~Y.}\ \bibnamefont
  {Illarionov}},\ and\ \bibinfo {author} {\bibfnamefont {V.}~\bibnamefont
  {Toneev}},\ }\href@noop {} {\bibfield  {journal} {\bibinfo  {journal}
  {International Journal of Modern Physics A}\ }\textbf {\bibinfo {volume}
  {24}},\ \bibinfo {pages} {5925} (\bibinfo {year} {2009})}\BibitemShut
  {NoStop}%
\bibitem [{\citenamefont {Deng}\ and\ \citenamefont
  {Huang}(2015)}]{deng2015electric}%
  \BibitemOpen
  \bibfield  {author} {\bibinfo {author} {\bibfnamefont {W.-T.}\ \bibnamefont
  {Deng}}\ and\ \bibinfo {author} {\bibfnamefont {X.-G.}\ \bibnamefont
  {Huang}},\ }\href@noop {} {\bibfield  {journal} {\bibinfo  {journal} {Physics
  Letters B}\ }\textbf {\bibinfo {volume} {742}},\ \bibinfo {pages} {296}
  (\bibinfo {year} {2015})}\BibitemShut {NoStop}%
\bibitem [{\citenamefont {Inghirami}\ \emph {et~al.}(2020)\citenamefont
  {Inghirami}, \citenamefont {Mace}, \citenamefont {Hirono}, \citenamefont
  {Del~Zanna}, \citenamefont {Kharzeev},\ and\ \citenamefont
  {Bleicher}}]{inghirami2020magnetic}%
  \BibitemOpen
  \bibfield  {author} {\bibinfo {author} {\bibfnamefont {G.}~\bibnamefont
  {Inghirami}}, \bibinfo {author} {\bibfnamefont {M.}~\bibnamefont {Mace}},
  \bibinfo {author} {\bibfnamefont {Y.}~\bibnamefont {Hirono}}, \bibinfo
  {author} {\bibfnamefont {L.}~\bibnamefont {Del~Zanna}}, \bibinfo {author}
  {\bibfnamefont {D.~E.}\ \bibnamefont {Kharzeev}},\ and\ \bibinfo {author}
  {\bibfnamefont {M.}~\bibnamefont {Bleicher}},\ }\href@noop {} {\bibfield
  {journal} {\bibinfo  {journal} {The European Physical Journal C}\ }\textbf
  {\bibinfo {volume} {80}},\ \bibinfo {pages} {293} (\bibinfo {year}
  {2020})}\BibitemShut {NoStop}%
\bibitem [{\citenamefont {Kharzeev}\ \emph {et~al.}(2008)\citenamefont
  {Kharzeev}, \citenamefont {McLerran},\ and\ \citenamefont
  {Warringa}}]{Kharzeev:2007jp}%
  \BibitemOpen
  \bibfield  {author} {\bibinfo {author} {\bibfnamefont {D.~E.}\ \bibnamefont
  {Kharzeev}}, \bibinfo {author} {\bibfnamefont {L.~D.}\ \bibnamefont
  {McLerran}},\ and\ \bibinfo {author} {\bibfnamefont {H.~J.}\ \bibnamefont
  {Warringa}},\ }\href {https://doi.org/10.1016/j.nuclphysa.2008.02.298}
  {\bibfield  {journal} {\bibinfo  {journal} {Nucl. Phys.}\ }\textbf {\bibinfo
  {volume} {A803}},\ \bibinfo {pages} {227} (\bibinfo {year} {2008})},\ \Eprint
  {https://arxiv.org/abs/0711.0950} {arXiv:0711.0950 [hep-ph]} \BibitemShut
  {NoStop}%
\bibitem [{\citenamefont {Abdulhamid}\ \emph {et~al.}(2024)\citenamefont
  {Abdulhamid}, \citenamefont {Aboona}, \citenamefont {Adam}, \citenamefont
  {Adams}, \citenamefont {Agakishiev}, \citenamefont {Aggarwal}, \citenamefont
  {Aggarwal}, \citenamefont {Ahammed}, \citenamefont {Aitbaev}, \citenamefont
  {Alekseev} \emph {et~al.}}]{abdulhamid2024observation}%
  \BibitemOpen
  \bibfield  {author} {\bibinfo {author} {\bibfnamefont {M.}~\bibnamefont
  {Abdulhamid}}, \bibinfo {author} {\bibfnamefont {B.}~\bibnamefont {Aboona}},
  \bibinfo {author} {\bibfnamefont {J.}~\bibnamefont {Adam}}, \bibinfo {author}
  {\bibfnamefont {J.}~\bibnamefont {Adams}}, \bibinfo {author} {\bibfnamefont
  {G.}~\bibnamefont {Agakishiev}}, \bibinfo {author} {\bibfnamefont
  {I.}~\bibnamefont {Aggarwal}}, \bibinfo {author} {\bibfnamefont
  {M.}~\bibnamefont {Aggarwal}}, \bibinfo {author} {\bibfnamefont
  {Z.}~\bibnamefont {Ahammed}}, \bibinfo {author} {\bibfnamefont
  {A.}~\bibnamefont {Aitbaev}}, \bibinfo {author} {\bibfnamefont
  {I.}~\bibnamefont {Alekseev}}, \emph {et~al.},\ }\href@noop {} {\bibfield
  {journal} {\bibinfo  {journal} {Physical Review X}\ }\textbf {\bibinfo
  {volume} {14}},\ \bibinfo {pages} {011028} (\bibinfo {year}
  {2024})}\BibitemShut {NoStop}%
\bibitem [{\citenamefont {Grasso}\ and\ \citenamefont
  {Rubinstein}(2001)}]{grasso2001magnetic}%
  \BibitemOpen
  \bibfield  {author} {\bibinfo {author} {\bibfnamefont {D.}~\bibnamefont
  {Grasso}}\ and\ \bibinfo {author} {\bibfnamefont {H.~R.}\ \bibnamefont
  {Rubinstein}},\ }\href@noop {} {\bibfield  {journal} {\bibinfo  {journal}
  {Physics Reports}\ }\textbf {\bibinfo {volume} {348}},\ \bibinfo {pages}
  {163} (\bibinfo {year} {2001})}\BibitemShut {NoStop}%
\bibitem [{\citenamefont {Vachaspati}(1991)}]{vachaspati1991magnetic}%
  \BibitemOpen
  \bibfield  {author} {\bibinfo {author} {\bibfnamefont {T.}~\bibnamefont
  {Vachaspati}},\ }\href@noop {} {\bibfield  {journal} {\bibinfo  {journal}
  {Physics Letters B}\ }\textbf {\bibinfo {volume} {265}},\ \bibinfo {pages}
  {258} (\bibinfo {year} {1991})}\BibitemShut {NoStop}%
\bibitem [{\citenamefont {Vachaspati}(2021)}]{vachaspati2021progress}%
  \BibitemOpen
  \bibfield  {author} {\bibinfo {author} {\bibfnamefont {T.}~\bibnamefont
  {Vachaspati}},\ }\href@noop {} {\bibfield  {journal} {\bibinfo  {journal}
  {Reports on progress in physics}\ }\textbf {\bibinfo {volume} {84}},\
  \bibinfo {pages} {074901} (\bibinfo {year} {2021})}\BibitemShut {NoStop}%
\bibitem [{\citenamefont {Enqvist}\ and\ \citenamefont
  {Olesen}(1993)}]{enqvist1993primordial}%
  \BibitemOpen
  \bibfield  {author} {\bibinfo {author} {\bibfnamefont {K.}~\bibnamefont
  {Enqvist}}\ and\ \bibinfo {author} {\bibfnamefont {P.}~\bibnamefont
  {Olesen}},\ }\href@noop {} {\bibfield  {journal} {\bibinfo  {journal}
  {Physics Letters B}\ }\textbf {\bibinfo {volume} {319}},\ \bibinfo {pages}
  {178} (\bibinfo {year} {1993})}\BibitemShut {NoStop}%
\bibitem [{\citenamefont {Baym}\ \emph {et~al.}(1996)\citenamefont {Baym},
  \citenamefont {B{\"o}deker},\ and\ \citenamefont
  {McLerran}}]{baym1996magnetic}%
  \BibitemOpen
  \bibfield  {author} {\bibinfo {author} {\bibfnamefont {G.}~\bibnamefont
  {Baym}}, \bibinfo {author} {\bibfnamefont {D.}~\bibnamefont {B{\"o}deker}},\
  and\ \bibinfo {author} {\bibfnamefont {L.}~\bibnamefont {McLerran}},\
  }\href@noop {} {\bibfield  {journal} {\bibinfo  {journal} {Physical Review
  D}\ }\textbf {\bibinfo {volume} {53}},\ \bibinfo {pages} {662} (\bibinfo
  {year} {1996})}\BibitemShut {NoStop}%
\bibitem [{\citenamefont {D'Elia}\ \emph {et~al.}(2010)\citenamefont {D'Elia},
  \citenamefont {Mukherjee},\ and\ \citenamefont {Sanfilippo}}]{d2010qcd}%
  \BibitemOpen
  \bibfield  {author} {\bibinfo {author} {\bibfnamefont {M.}~\bibnamefont
  {D'Elia}}, \bibinfo {author} {\bibfnamefont {S.}~\bibnamefont {Mukherjee}},\
  and\ \bibinfo {author} {\bibfnamefont {F.}~\bibnamefont {Sanfilippo}},\
  }\href@noop {} {\bibfield  {journal} {\bibinfo  {journal} {Physical Review
  D}\ }\textbf {\bibinfo {volume} {82}},\ \bibinfo {pages} {051501} (\bibinfo
  {year} {2010})}\BibitemShut {NoStop}%
\bibitem [{\citenamefont {D'Elia}\ \emph {et~al.}(2013)\citenamefont {D'Elia},
  \citenamefont {Mariti},\ and\ \citenamefont {Negro}}]{d2013susceptibility}%
  \BibitemOpen
  \bibfield  {author} {\bibinfo {author} {\bibfnamefont {M.}~\bibnamefont
  {D'Elia}}, \bibinfo {author} {\bibfnamefont {M.}~\bibnamefont {Mariti}},\
  and\ \bibinfo {author} {\bibfnamefont {F.}~\bibnamefont {Negro}},\
  }\href@noop {} {\bibfield  {journal} {\bibinfo  {journal} {Physical Review
  Letters}\ }\textbf {\bibinfo {volume} {110}},\ \bibinfo {pages} {082002}
  (\bibinfo {year} {2013})}\BibitemShut {NoStop}%
\bibitem [{\citenamefont {Bali}\ \emph {et~al.}(2014)\citenamefont {Bali},
  \citenamefont {Bruckmann}, \citenamefont {Endr{\H{o}}di}, \citenamefont
  {Katz},\ and\ \citenamefont {Sch{\"a}fer}}]{bali2014qcd}%
  \BibitemOpen
  \bibfield  {author} {\bibinfo {author} {\bibfnamefont {G.~S.}\ \bibnamefont
  {Bali}}, \bibinfo {author} {\bibfnamefont {F.}~\bibnamefont {Bruckmann}},
  \bibinfo {author} {\bibfnamefont {G.}~\bibnamefont {Endr{\H{o}}di}}, \bibinfo
  {author} {\bibfnamefont {S.}~\bibnamefont {Katz}},\ and\ \bibinfo {author}
  {\bibfnamefont {A.}~\bibnamefont {Sch{\"a}fer}},\ }\href@noop {} {\bibfield
  {journal} {\bibinfo  {journal} {Journal of High Energy Physics}\ }\textbf
  {\bibinfo {volume} {2014}},\ \bibinfo {pages} {1} (\bibinfo {year}
  {2014})}\BibitemShut {NoStop}%
\bibitem [{\citenamefont {Bali}\ \emph {et~al.}(2020)\citenamefont {Bali},
  \citenamefont {Endr{\H{o}}di},\ and\ \citenamefont
  {Piemonte}}]{bali2020magnetic}%
  \BibitemOpen
  \bibfield  {author} {\bibinfo {author} {\bibfnamefont {G.~S.}\ \bibnamefont
  {Bali}}, \bibinfo {author} {\bibfnamefont {G.}~\bibnamefont
  {Endr{\H{o}}di}},\ and\ \bibinfo {author} {\bibfnamefont {S.}~\bibnamefont
  {Piemonte}},\ }\href@noop {} {\bibfield  {journal} {\bibinfo  {journal}
  {Journal of High Energy Physics}\ }\textbf {\bibinfo {volume} {2020}},\
  \bibinfo {pages} {1} (\bibinfo {year} {2020})}\BibitemShut {NoStop}%
\bibitem [{\citenamefont {Bali}\ \emph {et~al.}(2013)\citenamefont {Bali},
  \citenamefont {Bruckmann}, \citenamefont {Endr{\H{o}}di}, \citenamefont
  {Gruber},\ and\ \citenamefont {Schaefer}}]{bali2013magnetic}%
  \BibitemOpen
  \bibfield  {author} {\bibinfo {author} {\bibfnamefont {G.}~\bibnamefont
  {Bali}}, \bibinfo {author} {\bibfnamefont {F.}~\bibnamefont {Bruckmann}},
  \bibinfo {author} {\bibfnamefont {G.}~\bibnamefont {Endr{\H{o}}di}}, \bibinfo
  {author} {\bibfnamefont {F.}~\bibnamefont {Gruber}},\ and\ \bibinfo {author}
  {\bibfnamefont {A.}~\bibnamefont {Schaefer}},\ }\href@noop {} {\bibfield
  {journal} {\bibinfo  {journal} {Journal of High Energy Physics}\ }\textbf
  {\bibinfo {volume} {2013}},\ \bibinfo {pages} {1} (\bibinfo {year}
  {2013})}\BibitemShut {NoStop}%
\bibitem [{\citenamefont {Bruckmann}\ \emph {et~al.}(2013)\citenamefont
  {Bruckmann}, \citenamefont {Endr{\H{o}}di},\ and\ \citenamefont
  {Kovacs}}]{bruckmann2013inverse}%
  \BibitemOpen
  \bibfield  {author} {\bibinfo {author} {\bibfnamefont {F.}~\bibnamefont
  {Bruckmann}}, \bibinfo {author} {\bibfnamefont {G.}~\bibnamefont
  {Endr{\H{o}}di}},\ and\ \bibinfo {author} {\bibfnamefont {T.~G.}\
  \bibnamefont {Kovacs}},\ }\href@noop {} {\bibfield  {journal} {\bibinfo
  {journal} {Journal of High Energy Physics}\ }\textbf {\bibinfo {volume}
  {2013}},\ \bibinfo {pages} {1} (\bibinfo {year} {2013})}\BibitemShut
  {NoStop}%
\bibitem [{\citenamefont {Bornyakov}\ \emph {et~al.}(2014)\citenamefont
  {Bornyakov}, \citenamefont {Buividovich}, \citenamefont {Cundy},
  \citenamefont {Kochetkov},\ and\ \citenamefont
  {Sch{\"a}fer}}]{bornyakov2014deconfinement}%
  \BibitemOpen
  \bibfield  {author} {\bibinfo {author} {\bibfnamefont {V.}~\bibnamefont
  {Bornyakov}}, \bibinfo {author} {\bibfnamefont {P.}~\bibnamefont
  {Buividovich}}, \bibinfo {author} {\bibfnamefont {N.}~\bibnamefont {Cundy}},
  \bibinfo {author} {\bibfnamefont {O.}~\bibnamefont {Kochetkov}},\ and\
  \bibinfo {author} {\bibfnamefont {A.}~\bibnamefont {Sch{\"a}fer}},\
  }\href@noop {} {\bibfield  {journal} {\bibinfo  {journal} {Physical Review
  D}\ }\textbf {\bibinfo {volume} {90}},\ \bibinfo {pages} {034501} (\bibinfo
  {year} {2014})}\BibitemShut {NoStop}%
\bibitem [{\citenamefont {Endr{\"o}di}(2015)}]{endrodi2015critical}%
  \BibitemOpen
  \bibfield  {author} {\bibinfo {author} {\bibfnamefont {G.}~\bibnamefont
  {Endr{\"o}di}},\ }\href@noop {} {\bibfield  {journal} {\bibinfo  {journal}
  {Journal of High Energy Physics}\ }\textbf {\bibinfo {volume} {2015}},\
  \bibinfo {pages} {1} (\bibinfo {year} {2015})}\BibitemShut {NoStop}%
\bibitem [{\citenamefont {Ding}\ \emph {et~al.}(2020)\citenamefont {Ding},
  \citenamefont {Li}, \citenamefont {Shi}, \citenamefont {Tomiya},
  \citenamefont {Wang},\ and\ \citenamefont {Zhang}}]{ding2020qcd}%
  \BibitemOpen
  \bibfield  {author} {\bibinfo {author} {\bibfnamefont {H.-T.}\ \bibnamefont
  {Ding}}, \bibinfo {author} {\bibfnamefont {S.-T.}\ \bibnamefont {Li}},
  \bibinfo {author} {\bibfnamefont {Q.}~\bibnamefont {Shi}}, \bibinfo {author}
  {\bibfnamefont {A.}~\bibnamefont {Tomiya}}, \bibinfo {author} {\bibfnamefont
  {X.-D.}\ \bibnamefont {Wang}},\ and\ \bibinfo {author} {\bibfnamefont
  {Y.}~\bibnamefont {Zhang}},\ }\href@noop {} {\bibfield  {journal} {\bibinfo
  {journal} {arXiv preprint arXiv:2011.04870}\ } (\bibinfo {year}
  {2020})}\BibitemShut {NoStop}%
\bibitem [{\citenamefont {D'Elia}\ \emph {et~al.}(2021)\citenamefont {D'Elia},
  \citenamefont {Maio}, \citenamefont {Sanfilippo},\ and\ \citenamefont
  {Stanzione}}]{d2021confining}%
  \BibitemOpen
  \bibfield  {author} {\bibinfo {author} {\bibfnamefont {M.}~\bibnamefont
  {D'Elia}}, \bibinfo {author} {\bibfnamefont {L.}~\bibnamefont {Maio}},
  \bibinfo {author} {\bibfnamefont {F.}~\bibnamefont {Sanfilippo}},\ and\
  \bibinfo {author} {\bibfnamefont {A.}~\bibnamefont {Stanzione}},\ }\href@noop
  {} {\bibfield  {journal} {\bibinfo  {journal} {Physical Review D}\ }\textbf
  {\bibinfo {volume} {104}},\ \bibinfo {pages} {114512} (\bibinfo {year}
  {2021})}\BibitemShut {NoStop}%
\bibitem [{\citenamefont {Ding}\ \emph {et~al.}(2022)\citenamefont {Ding},
  \citenamefont {Li}, \citenamefont {Liu},\ and\ \citenamefont
  {Wang}}]{ding2022chiral}%
  \BibitemOpen
  \bibfield  {author} {\bibinfo {author} {\bibfnamefont {H.-T.}\ \bibnamefont
  {Ding}}, \bibinfo {author} {\bibfnamefont {S.-T.}\ \bibnamefont {Li}},
  \bibinfo {author} {\bibfnamefont {J.-H.}\ \bibnamefont {Liu}},\ and\ \bibinfo
  {author} {\bibfnamefont {X.-D.}\ \bibnamefont {Wang}},\ }\href@noop {}
  {\bibfield  {journal} {\bibinfo  {journal} {Physical Review D}\ }\textbf
  {\bibinfo {volume} {105}},\ \bibinfo {pages} {034514} (\bibinfo {year}
  {2022})}\BibitemShut {NoStop}%
\bibitem [{\citenamefont {Andersen}\ \emph {et~al.}(2016)\citenamefont
  {Andersen}, \citenamefont {Naylor},\ and\ \citenamefont
  {Tranberg}}]{Andersen:2014xxa}%
  \BibitemOpen
  \bibfield  {author} {\bibinfo {author} {\bibfnamefont {J.~O.}\ \bibnamefont
  {Andersen}}, \bibinfo {author} {\bibfnamefont {W.~R.}\ \bibnamefont
  {Naylor}},\ and\ \bibinfo {author} {\bibfnamefont {A.}~\bibnamefont
  {Tranberg}},\ }\href {https://doi.org/10.1103/RevModPhys.88.025001}
  {\bibfield  {journal} {\bibinfo  {journal} {Rev. Mod. Phys.}\ }\textbf
  {\bibinfo {volume} {88}},\ \bibinfo {pages} {025001} (\bibinfo {year}
  {2016})},\ \Eprint {https://arxiv.org/abs/1411.7176} {arXiv:1411.7176
  [hep-ph]} \BibitemShut {NoStop}%
\bibitem [{\citenamefont {Shovkovy}(2013)}]{shovkovy2013magnetic}%
  \BibitemOpen
  \bibfield  {author} {\bibinfo {author} {\bibfnamefont {I.~A.}\ \bibnamefont
  {Shovkovy}},\ }\href@noop {} {\bibfield  {journal} {\bibinfo  {journal}
  {Strongly Interacting Matter in Magnetic Fields}\ ,\ \bibinfo {pages} {13}}
  (\bibinfo {year} {2013})}\BibitemShut {NoStop}%
\bibitem [{\citenamefont {Kharzeev}\ \emph {et~al.}(2013)\citenamefont
  {Kharzeev}, \citenamefont {Landsteiner}, \citenamefont {Schmitt},\ and\
  \citenamefont {Yee}}]{kharzeev2013strongly}%
  \BibitemOpen
  \bibfield  {author} {\bibinfo {author} {\bibfnamefont {D.~E.}\ \bibnamefont
  {Kharzeev}}, \bibinfo {author} {\bibfnamefont {K.}~\bibnamefont
  {Landsteiner}}, \bibinfo {author} {\bibfnamefont {A.}~\bibnamefont
  {Schmitt}},\ and\ \bibinfo {author} {\bibfnamefont {H.-U.}\ \bibnamefont
  {Yee}},\ }\href@noop {} {\emph {\bibinfo {title} {Strongly interacting matter
  in magnetic fields: a guide to this volume}}}\ (\bibinfo  {publisher}
  {Springer},\ \bibinfo {year} {2013})\BibitemShut {NoStop}%
\bibitem [{\citenamefont {Endr{\H{o}}di}(2025)}]{endrHodi2025qcd}%
  \BibitemOpen
  \bibfield  {author} {\bibinfo {author} {\bibfnamefont {G.}~\bibnamefont
  {Endr{\H{o}}di}},\ }\href@noop {} {\bibfield  {journal} {\bibinfo  {journal}
  {Progress in Particle and Nuclear Physics}\ }\textbf {\bibinfo {volume}
  {141}},\ \bibinfo {pages} {104153} (\bibinfo {year} {2025})}\BibitemShut
  {NoStop}%
\bibitem [{\citenamefont {Adhikari}\ \emph {et~al.}(2026)\citenamefont
  {Adhikari}, \citenamefont {Ammon}, \citenamefont {Avancini}, \citenamefont
  {Ayala}, \citenamefont {Bandyopadhyay}, \citenamefont {Blaschke},
  \citenamefont {Braghin}, \citenamefont {Buividovich}, \citenamefont
  {Cardoso}, \citenamefont {Cartwright} \emph {et~al.}}]{adhikari2025strongly}%
  \BibitemOpen
  \bibfield  {author} {\bibinfo {author} {\bibfnamefont {P.}~\bibnamefont
  {Adhikari}}, \bibinfo {author} {\bibfnamefont {M.}~\bibnamefont {Ammon}},
  \bibinfo {author} {\bibfnamefont {S.~S.}\ \bibnamefont {Avancini}}, \bibinfo
  {author} {\bibfnamefont {A.}~\bibnamefont {Ayala}}, \bibinfo {author}
  {\bibfnamefont {A.}~\bibnamefont {Bandyopadhyay}}, \bibinfo {author}
  {\bibfnamefont {D.}~\bibnamefont {Blaschke}}, \bibinfo {author}
  {\bibfnamefont {F.~L.}\ \bibnamefont {Braghin}}, \bibinfo {author}
  {\bibfnamefont {P.}~\bibnamefont {Buividovich}}, \bibinfo {author}
  {\bibfnamefont {R.~P.}\ \bibnamefont {Cardoso}}, \bibinfo {author}
  {\bibfnamefont {C.}~\bibnamefont {Cartwright}}, \emph {et~al.},\ }\href@noop
  {} {\bibfield  {journal} {\bibinfo  {journal} {Progress in Particle and
  Nuclear Physics}\ }\textbf {\bibinfo {volume} {146}},\ \bibinfo {pages}
  {104199} (\bibinfo {year} {2026})}\BibitemShut {NoStop}%
\bibitem [{\citenamefont {Shushpanov}\ and\ \citenamefont
  {Smilga}(1997)}]{shushpanov1997quark}%
  \BibitemOpen
  \bibfield  {author} {\bibinfo {author} {\bibfnamefont {I.}~\bibnamefont
  {Shushpanov}}\ and\ \bibinfo {author} {\bibfnamefont {A.~V.}\ \bibnamefont
  {Smilga}},\ }\href@noop {} {\bibfield  {journal} {\bibinfo  {journal}
  {Physics Letters B}\ }\textbf {\bibinfo {volume} {402}},\ \bibinfo {pages}
  {351} (\bibinfo {year} {1997})}\BibitemShut {NoStop}%
\bibitem [{\citenamefont {Agasian}\ and\ \citenamefont
  {Shushpanov}(2000)}]{agasian2000quark}%
  \BibitemOpen
  \bibfield  {author} {\bibinfo {author} {\bibfnamefont {N.~O.}\ \bibnamefont
  {Agasian}}\ and\ \bibinfo {author} {\bibfnamefont {I.}~\bibnamefont
  {Shushpanov}},\ }\href@noop {} {\bibfield  {journal} {\bibinfo  {journal}
  {Physics Letters B}\ }\textbf {\bibinfo {volume} {472}},\ \bibinfo {pages}
  {143} (\bibinfo {year} {2000})}\BibitemShut {NoStop}%
\bibitem [{\citenamefont {Cohen}\ \emph {et~al.}(2007)\citenamefont {Cohen},
  \citenamefont {McGady},\ and\ \citenamefont {Werbos}}]{Cohen:2007bt}%
  \BibitemOpen
  \bibfield  {author} {\bibinfo {author} {\bibfnamefont {T.~D.}\ \bibnamefont
  {Cohen}}, \bibinfo {author} {\bibfnamefont {D.~A.}\ \bibnamefont {McGady}},\
  and\ \bibinfo {author} {\bibfnamefont {E.~S.}\ \bibnamefont {Werbos}},\
  }\href {https://doi.org/10.1103/PhysRevC.76.055201} {\bibfield  {journal}
  {\bibinfo  {journal} {Phys. Rev.}\ }\textbf {\bibinfo {volume} {C76}},\
  \bibinfo {pages} {055201} (\bibinfo {year} {2007})},\ \Eprint
  {https://arxiv.org/abs/0706.3208} {arXiv:0706.3208 [hep-ph]} \BibitemShut
  {NoStop}%
\bibitem [{\citenamefont {Werbos}(2008)}]{Werbos:2007ym}%
  \BibitemOpen
  \bibfield  {author} {\bibinfo {author} {\bibfnamefont {E.~S.}\ \bibnamefont
  {Werbos}},\ }\href {https://doi.org/10.1103/PhysRevC.77.065202} {\bibfield
  {journal} {\bibinfo  {journal} {Phys. Rev.}\ }\textbf {\bibinfo {volume}
  {C77}},\ \bibinfo {pages} {065202} (\bibinfo {year} {2008})},\ \Eprint
  {https://arxiv.org/abs/0711.2635} {arXiv:0711.2635 [hep-ph]} \BibitemShut
  {NoStop}%
\bibitem [{\citenamefont {Hofmann}(2020)}]{hofmann2020chiral}%
  \BibitemOpen
  \bibfield  {author} {\bibinfo {author} {\bibfnamefont {C.~P.}\ \bibnamefont
  {Hofmann}},\ }\href@noop {} {\bibfield  {journal} {\bibinfo  {journal}
  {Physical Review D}\ }\textbf {\bibinfo {volume} {102}},\ \bibinfo {pages}
  {094010} (\bibinfo {year} {2020})}\BibitemShut {NoStop}%
\bibitem [{\citenamefont
  {Hofmann}(2021{\natexlab{a}})}]{hofmann2021diamagnetic}%
  \BibitemOpen
  \bibfield  {author} {\bibinfo {author} {\bibfnamefont {C.~P.}\ \bibnamefont
  {Hofmann}},\ }\href@noop {} {\bibfield  {journal} {\bibinfo  {journal}
  {Physics Letters B}\ }\textbf {\bibinfo {volume} {818}},\ \bibinfo {pages}
  {136384} (\bibinfo {year} {2021}{\natexlab{a}})}\BibitemShut {NoStop}%
\bibitem [{\citenamefont
  {Hofmann}(2021{\natexlab{b}})}]{hofmann2021thermomagnetic}%
  \BibitemOpen
  \bibfield  {author} {\bibinfo {author} {\bibfnamefont {C.~P.}\ \bibnamefont
  {Hofmann}},\ }\href@noop {} {\bibfield  {journal} {\bibinfo  {journal}
  {Physical Review D}\ }\textbf {\bibinfo {volume} {104}},\ \bibinfo {pages}
  {014025} (\bibinfo {year} {2021}{\natexlab{b}})}\BibitemShut {NoStop}%
\bibitem [{\citenamefont {Adhikari}(2022)}]{Adhikari:2021lbl}%
  \BibitemOpen
  \bibfield  {author} {\bibinfo {author} {\bibfnamefont {P.}~\bibnamefont
  {Adhikari}},\ }\href@noop {} {\bibfield  {journal} {\bibinfo  {journal}
  {Phys. Lett. B}\ }\textbf {\bibinfo {volume} {825}} (\bibinfo {year}
  {2022})},\ \Eprint {https://arxiv.org/abs/2103.05048} {arXiv:2103.05048
  [hep-ph]} \BibitemShut {NoStop}%
\bibitem [{\citenamefont {Cohen}\ and\ \citenamefont
  {Yamamoto}(2014)}]{cohen2014new}%
  \BibitemOpen
  \bibfield  {author} {\bibinfo {author} {\bibfnamefont {T.~D.}\ \bibnamefont
  {Cohen}}\ and\ \bibinfo {author} {\bibfnamefont {N.}~\bibnamefont
  {Yamamoto}},\ }\href@noop {} {\bibfield  {journal} {\bibinfo  {journal}
  {Physical Review D}\ }\textbf {\bibinfo {volume} {89}},\ \bibinfo {pages}
  {054029} (\bibinfo {year} {2014})}\BibitemShut {NoStop}%
\bibitem [{\citenamefont {Adhikari}\ \emph {et~al.}(2015)\citenamefont
  {Adhikari}, \citenamefont {Cohen},\ and\ \citenamefont
  {Sakowitz}}]{Adhikari:2015wva}%
  \BibitemOpen
  \bibfield  {author} {\bibinfo {author} {\bibfnamefont {P.}~\bibnamefont
  {Adhikari}}, \bibinfo {author} {\bibfnamefont {T.~D.}\ \bibnamefont
  {Cohen}},\ and\ \bibinfo {author} {\bibfnamefont {J.}~\bibnamefont
  {Sakowitz}},\ }\href {https://doi.org/10.1103/PhysRevC.91.045202} {\bibfield
  {journal} {\bibinfo  {journal} {Phys. Rev.}\ }\textbf {\bibinfo {volume}
  {C91}},\ \bibinfo {pages} {045202} (\bibinfo {year} {2015})},\ \Eprint
  {https://arxiv.org/abs/1501.02737} {arXiv:1501.02737 [nucl-th]} \BibitemShut
  {NoStop}%
\bibitem [{\citenamefont {Adhikari}(2019)}]{Adhikari:2018fwm}%
  \BibitemOpen
  \bibfield  {author} {\bibinfo {author} {\bibfnamefont {P.}~\bibnamefont
  {Adhikari}},\ }\href {https://doi.org/10.1016/j.physletb.2019.01.027}
  {\bibfield  {journal} {\bibinfo  {journal} {Phys. Lett.}\ }\textbf {\bibinfo
  {volume} {B790}},\ \bibinfo {pages} {211} (\bibinfo {year} {2019})},\ \Eprint
  {https://arxiv.org/abs/1810.03663} {arXiv:1810.03663 [nucl-th]} \BibitemShut
  {NoStop}%
\bibitem [{\citenamefont {Adhikari}\ \emph {et~al.}(2022)\citenamefont
  {Adhikari}, \citenamefont {Leeser},\ and\ \citenamefont
  {Markowski}}]{adhikari2022phonon}%
  \BibitemOpen
  \bibfield  {author} {\bibinfo {author} {\bibfnamefont {P.}~\bibnamefont
  {Adhikari}}, \bibinfo {author} {\bibfnamefont {E.}~\bibnamefont {Leeser}},\
  and\ \bibinfo {author} {\bibfnamefont {J.}~\bibnamefont {Markowski}},\
  }\href@noop {} {\bibfield  {journal} {\bibinfo  {journal} {arXiv preprint
  arXiv:2205.13369}\ } (\bibinfo {year} {2022})}\BibitemShut {NoStop}%
\bibitem [{\citenamefont {Brauner}\ and\ \citenamefont
  {Yamamoto}(2017)}]{Brauner:2016pko}%
  \BibitemOpen
  \bibfield  {author} {\bibinfo {author} {\bibfnamefont {T.}~\bibnamefont
  {Brauner}}\ and\ \bibinfo {author} {\bibfnamefont {N.}~\bibnamefont
  {Yamamoto}},\ }\href {https://doi.org/10.1007/JHEP04(2017)132} {\bibfield
  {journal} {\bibinfo  {journal} {JHEP}\ }\textbf {\bibinfo {volume} {04}},\
  \bibinfo {pages} {132}},\ \Eprint {https://arxiv.org/abs/1609.05213}
  {arXiv:1609.05213 [hep-ph]} \BibitemShut {NoStop}%
\bibitem [{\citenamefont {Brauner}\ \emph {et~al.}(2021)\citenamefont
  {Brauner}, \citenamefont {Kole\v{s}ov\'a},\ and\ \citenamefont
  {Yamamoto}}]{Brauner:2021sci}%
  \BibitemOpen
  \bibfield  {author} {\bibinfo {author} {\bibfnamefont {T.}~\bibnamefont
  {Brauner}}, \bibinfo {author} {\bibfnamefont {H.}~\bibnamefont
  {Kole\v{s}ov\'a}},\ and\ \bibinfo {author} {\bibfnamefont {N.}~\bibnamefont
  {Yamamoto}},\ }\href {https://doi.org/10.1016/j.physletb.2021.136767}
  {\bibfield  {journal} {\bibinfo  {journal} {Phys. Lett. B}\ }\textbf
  {\bibinfo {volume} {823}},\ \bibinfo {pages} {136767} (\bibinfo {year}
  {2021})},\ \Eprint {https://arxiv.org/abs/2108.10044} {arXiv:2108.10044
  [hep-ph]} \BibitemShut {NoStop}%
\bibitem [{\citenamefont {Evans}\ and\ \citenamefont
  {Schmitt}(2022)}]{Evans:2022hwr}%
  \BibitemOpen
  \bibfield  {author} {\bibinfo {author} {\bibfnamefont {G.~W.}\ \bibnamefont
  {Evans}}\ and\ \bibinfo {author} {\bibfnamefont {A.}~\bibnamefont
  {Schmitt}},\ }\href {https://doi.org/10.1007/JHEP09(2022)192} {\bibfield
  {journal} {\bibinfo  {journal} {JHEP}\ }\textbf {\bibinfo {volume} {09}},\
  \bibinfo {pages} {192}},\ \Eprint {https://arxiv.org/abs/2206.01227}
  {arXiv:2206.01227 [hep-th]} \BibitemShut {NoStop}%
\bibitem [{\citenamefont {Hamada}\ \emph {et~al.}(2026)\citenamefont {Hamada},
  \citenamefont {Nitta},\ and\ \citenamefont {Qiu}}]{hamada2026qcd}%
  \BibitemOpen
  \bibfield  {author} {\bibinfo {author} {\bibfnamefont {Y.}~\bibnamefont
  {Hamada}}, \bibinfo {author} {\bibfnamefont {M.}~\bibnamefont {Nitta}},\ and\
  \bibinfo {author} {\bibfnamefont {Z.}~\bibnamefont {Qiu}},\ }\href@noop {}
  {\bibfield  {journal} {\bibinfo  {journal} {arXiv preprint arXiv:2602.11762}\
  } (\bibinfo {year} {2026})}\BibitemShut {NoStop}%
\bibitem [{\citenamefont {Gasser}\ and\ \citenamefont
  {Leutwyler}(1984)}]{Gasser:1983yg}%
  \BibitemOpen
  \bibfield  {author} {\bibinfo {author} {\bibfnamefont {J.}~\bibnamefont
  {Gasser}}\ and\ \bibinfo {author} {\bibfnamefont {H.}~\bibnamefont
  {Leutwyler}},\ }\href {https://doi.org/10.1016/0003-4916(84)90242-2}
  {\bibfield  {journal} {\bibinfo  {journal} {Annals Phys.}\ }\textbf {\bibinfo
  {volume} {158}},\ \bibinfo {pages} {142} (\bibinfo {year}
  {1984})}\BibitemShut {NoStop}%
\bibitem [{\citenamefont {Gasser}\ and\ \citenamefont
  {Leutwyler}(1985)}]{Gasser:1984gg}%
  \BibitemOpen
  \bibfield  {author} {\bibinfo {author} {\bibfnamefont {J.}~\bibnamefont
  {Gasser}}\ and\ \bibinfo {author} {\bibfnamefont {H.}~\bibnamefont
  {Leutwyler}},\ }\href {https://doi.org/10.1016/0550-3213(85)90492-4}
  {\bibfield  {journal} {\bibinfo  {journal} {Nucl. Phys.}\ }\textbf {\bibinfo
  {volume} {B250}},\ \bibinfo {pages} {465} (\bibinfo {year}
  {1985})}\BibitemShut {NoStop}%
\bibitem [{\citenamefont {Gasser}\ and\ \citenamefont
  {Leutwyler}(1988)}]{gasser1988spontaneously}%
  \BibitemOpen
  \bibfield  {author} {\bibinfo {author} {\bibfnamefont {J.}~\bibnamefont
  {Gasser}}\ and\ \bibinfo {author} {\bibfnamefont {H.}~\bibnamefont
  {Leutwyler}},\ }\href@noop {} {\bibfield  {journal} {\bibinfo  {journal}
  {Nuclear Physics B}\ }\textbf {\bibinfo {volume} {307}},\ \bibinfo {pages}
  {763} (\bibinfo {year} {1988})}\BibitemShut {NoStop}%
\bibitem [{\citenamefont {Adhikari}\ and\ \citenamefont
  {Tiburzi}(2023)}]{adhikari2023qcd}%
  \BibitemOpen
  \bibfield  {author} {\bibinfo {author} {\bibfnamefont {P.}~\bibnamefont
  {Adhikari}}\ and\ \bibinfo {author} {\bibfnamefont {B.~C.}\ \bibnamefont
  {Tiburzi}},\ }\href@noop {} {\bibfield  {journal} {\bibinfo  {journal}
  {Physical Review D}\ }\textbf {\bibinfo {volume} {107}},\ \bibinfo {pages}
  {094504} (\bibinfo {year} {2023})}\BibitemShut {NoStop}%
\bibitem [{\citenamefont {Tiburzi}(2008)}]{tiburzi2008hadrons}%
  \BibitemOpen
  \bibfield  {author} {\bibinfo {author} {\bibfnamefont {B.~C.}\ \bibnamefont
  {Tiburzi}},\ }\href@noop {} {\bibfield  {journal} {\bibinfo  {journal}
  {Nuclear Physics A}\ }\textbf {\bibinfo {volume} {814}},\ \bibinfo {pages}
  {74} (\bibinfo {year} {2008})}\BibitemShut {NoStop}%
\bibitem [{\citenamefont {Andersen}(2012{\natexlab{a}})}]{andersen2012chiral}%
  \BibitemOpen
  \bibfield  {author} {\bibinfo {author} {\bibfnamefont {J.~O.}\ \bibnamefont
  {Andersen}},\ }\href@noop {} {\bibfield  {journal} {\bibinfo  {journal}
  {Journal of High Energy Physics}\ }\textbf {\bibinfo {volume} {2012}},\
  \bibinfo {pages} {1} (\bibinfo {year} {2012}{\natexlab{a}})}\BibitemShut
  {NoStop}%
\bibitem [{\citenamefont {Andersen}(2012{\natexlab{b}})}]{andersen2012thermal}%
  \BibitemOpen
  \bibfield  {author} {\bibinfo {author} {\bibfnamefont {J.~O.}\ \bibnamefont
  {Andersen}},\ }\href@noop {} {\bibfield  {journal} {\bibinfo  {journal}
  {Physical Review D}\ }\textbf {\bibinfo {volume} {86}},\ \bibinfo {pages}
  {025020} (\bibinfo {year} {2012}{\natexlab{b}})}\BibitemShut {NoStop}%
\bibitem [{\citenamefont {Adhikari}\ and\ \citenamefont
  {Tiburzi}(2024)}]{adhikari2024chiral}%
  \BibitemOpen
  \bibfield  {author} {\bibinfo {author} {\bibfnamefont {P.}~\bibnamefont
  {Adhikari}}\ and\ \bibinfo {author} {\bibfnamefont {B.~C.}\ \bibnamefont
  {Tiburzi}},\ }\href@noop {} {\bibfield  {journal} {\bibinfo  {journal} {arXiv
  preprint arXiv:2406.00818}\ } (\bibinfo {year} {2024})}\BibitemShut {NoStop}%
\bibitem [{\citenamefont {Bali}\ \emph
  {et~al.}(2018{\natexlab{a}})\citenamefont {Bali}, \citenamefont {Brandt},
  \citenamefont {Endr{\H{o}}di},\ and\ \citenamefont
  {Gl{\"a}{\ss}le}}]{bali2018weak}%
  \BibitemOpen
  \bibfield  {author} {\bibinfo {author} {\bibfnamefont {G.~S.}\ \bibnamefont
  {Bali}}, \bibinfo {author} {\bibfnamefont {B.~B.}\ \bibnamefont {Brandt}},
  \bibinfo {author} {\bibfnamefont {G.}~\bibnamefont {Endr{\H{o}}di}},\ and\
  \bibinfo {author} {\bibfnamefont {B.}~\bibnamefont {Gl{\"a}{\ss}le}},\
  }\href@noop {} {\bibfield  {journal} {\bibinfo  {journal} {Physical Review
  Letters}\ }\textbf {\bibinfo {volume} {121}},\ \bibinfo {pages} {072001}
  (\bibinfo {year} {2018}{\natexlab{a}})}\BibitemShut {NoStop}%
\bibitem [{\citenamefont {Coppola}\ \emph
  {et~al.}(2020{\natexlab{a}})\citenamefont {Coppola}, \citenamefont {Dumm},
  \citenamefont {Noguera},\ and\ \citenamefont {Scoccola}}]{coppola2020weak}%
  \BibitemOpen
  \bibfield  {author} {\bibinfo {author} {\bibfnamefont {M.}~\bibnamefont
  {Coppola}}, \bibinfo {author} {\bibfnamefont {D.~G.}\ \bibnamefont {Dumm}},
  \bibinfo {author} {\bibfnamefont {S.}~\bibnamefont {Noguera}},\ and\ \bibinfo
  {author} {\bibfnamefont {N.~N.}\ \bibnamefont {Scoccola}},\ }\href@noop {}
  {\bibfield  {journal} {\bibinfo  {journal} {Physical Review D}\ }\textbf
  {\bibinfo {volume} {101}},\ \bibinfo {pages} {034003} (\bibinfo {year}
  {2020}{\natexlab{a}})}\BibitemShut {NoStop}%
\bibitem [{\citenamefont {Coppola}\ \emph {et~al.}(2019)\citenamefont
  {Coppola}, \citenamefont {Dumm}, \citenamefont {Noguera},\ and\ \citenamefont
  {Scoccola}}]{coppola2019pion}%
  \BibitemOpen
  \bibfield  {author} {\bibinfo {author} {\bibfnamefont {M.}~\bibnamefont
  {Coppola}}, \bibinfo {author} {\bibfnamefont {D.~G.}\ \bibnamefont {Dumm}},
  \bibinfo {author} {\bibfnamefont {S.}~\bibnamefont {Noguera}},\ and\ \bibinfo
  {author} {\bibfnamefont {N.}~\bibnamefont {Scoccola}},\ }\href@noop {}
  {\bibfield  {journal} {\bibinfo  {journal} {Physical Review D}\ }\textbf
  {\bibinfo {volume} {99}},\ \bibinfo {pages} {054031} (\bibinfo {year}
  {2019})}\BibitemShut {NoStop}%
\bibitem [{\citenamefont {Coppola}\ \emph {et~al.}(2025)\citenamefont
  {Coppola}, \citenamefont {Gomez~Dumm},\ and\ \citenamefont
  {Scoccola}}]{coppola2025pi}%
  \BibitemOpen
  \bibfield  {author} {\bibinfo {author} {\bibfnamefont {M.}~\bibnamefont
  {Coppola}}, \bibinfo {author} {\bibfnamefont {D.}~\bibnamefont
  {Gomez~Dumm}},\ and\ \bibinfo {author} {\bibfnamefont {N.~N.}\ \bibnamefont
  {Scoccola}},\ }\href@noop {} {\bibfield  {journal} {\bibinfo  {journal}
  {Physical Review D}\ }\textbf {\bibinfo {volume} {112}},\ \bibinfo {pages}
  {054043} (\bibinfo {year} {2025})}\BibitemShut {NoStop}%
\bibitem [{\citenamefont {Deshmukh}\ and\ \citenamefont
  {Tiburzi}(2018)}]{deshmukh2018octet}%
  \BibitemOpen
  \bibfield  {author} {\bibinfo {author} {\bibfnamefont {A.}~\bibnamefont
  {Deshmukh}}\ and\ \bibinfo {author} {\bibfnamefont {B.~C.}\ \bibnamefont
  {Tiburzi}},\ }\href@noop {} {\bibfield  {journal} {\bibinfo  {journal}
  {Physical Review D}\ }\textbf {\bibinfo {volume} {97}},\ \bibinfo {pages}
  {014006} (\bibinfo {year} {2018})}\BibitemShut {NoStop}%
\bibitem [{\citenamefont {Ding}\ and\ \citenamefont
  {Zhang}(2026)}]{ding2026chiral}%
  \BibitemOpen
  \bibfield  {author} {\bibinfo {author} {\bibfnamefont {H.-T.}\ \bibnamefont
  {Ding}}\ and\ \bibinfo {author} {\bibfnamefont {D.}~\bibnamefont {Zhang}},\
  }\href@noop {} {\bibfield  {journal} {\bibinfo  {journal} {arXiv preprint
  arXiv:2601.18354}\ } (\bibinfo {year} {2026})}\BibitemShut {NoStop}%
\bibitem [{\citenamefont {Adhikari}\ and\ \citenamefont
  {Str{\"u}mke}(2023)}]{adhikari2023vacuum}%
  \BibitemOpen
  \bibfield  {author} {\bibinfo {author} {\bibfnamefont {P.}~\bibnamefont
  {Adhikari}}\ and\ \bibinfo {author} {\bibfnamefont {I.}~\bibnamefont
  {Str{\"u}mke}},\ }\href@noop {} {\bibfield  {journal} {\bibinfo  {journal}
  {Nuclear Physics B}\ }\textbf {\bibinfo {volume} {997}},\ \bibinfo {pages}
  {116389} (\bibinfo {year} {2023})}\BibitemShut {NoStop}%
\bibitem [{\citenamefont {Bali}\ \emph
  {et~al.}(2018{\natexlab{b}})\citenamefont {Bali}, \citenamefont {Brandt},
  \citenamefont {Endr{\H{o}}di},\ and\ \citenamefont
  {Gl{\"a}{\ss}le}}]{bali2018meson}%
  \BibitemOpen
  \bibfield  {author} {\bibinfo {author} {\bibfnamefont {G.~S.}\ \bibnamefont
  {Bali}}, \bibinfo {author} {\bibfnamefont {B.~B.}\ \bibnamefont {Brandt}},
  \bibinfo {author} {\bibfnamefont {G.}~\bibnamefont {Endr{\H{o}}di}},\ and\
  \bibinfo {author} {\bibfnamefont {B.}~\bibnamefont {Gl{\"a}{\ss}le}},\
  }\href@noop {} {\bibfield  {journal} {\bibinfo  {journal} {Physical Review
  D}\ }\textbf {\bibinfo {volume} {97}},\ \bibinfo {pages} {034505} (\bibinfo
  {year} {2018}{\natexlab{b}})}\BibitemShut {NoStop}%
\bibitem [{\citenamefont {Fayazbakhsh}\ \emph {et~al.}(2012)\citenamefont
  {Fayazbakhsh}, \citenamefont {Sadeghian},\ and\ \citenamefont
  {Sadooghi}}]{fayazbakhsh2012properties}%
  \BibitemOpen
  \bibfield  {author} {\bibinfo {author} {\bibfnamefont {S.}~\bibnamefont
  {Fayazbakhsh}}, \bibinfo {author} {\bibfnamefont {S.}~\bibnamefont
  {Sadeghian}},\ and\ \bibinfo {author} {\bibfnamefont {N.}~\bibnamefont
  {Sadooghi}},\ }\href@noop {} {\bibfield  {journal} {\bibinfo  {journal}
  {Physical Review D---Particles, Fields, Gravitation, and Cosmology}\ }\textbf
  {\bibinfo {volume} {86}},\ \bibinfo {pages} {085042} (\bibinfo {year}
  {2012})}\BibitemShut {NoStop}%
\bibitem [{\citenamefont {Abreu}\ \emph {et~al.}(2022)\citenamefont {Abreu},
  \citenamefont {Nery},\ and\ \citenamefont
  {Corr{\^e}a}}]{abreu2022properties}%
  \BibitemOpen
  \bibfield  {author} {\bibinfo {author} {\bibfnamefont {L.~M.}\ \bibnamefont
  {Abreu}}, \bibinfo {author} {\bibfnamefont {E.~S.}\ \bibnamefont {Nery}},\
  and\ \bibinfo {author} {\bibfnamefont {E.~B.}\ \bibnamefont {Corr{\^e}a}},\
  }\href@noop {} {\bibfield  {journal} {\bibinfo  {journal} {Physical Review
  D}\ }\textbf {\bibinfo {volume} {105}},\ \bibinfo {pages} {056010} (\bibinfo
  {year} {2022})}\BibitemShut {NoStop}%
\bibitem [{\citenamefont {Simonov}(2016)}]{simonov2016pion}%
  \BibitemOpen
  \bibfield  {author} {\bibinfo {author} {\bibfnamefont {Y.~A.}\ \bibnamefont
  {Simonov}},\ }\href@noop {} {\bibfield  {journal} {\bibinfo  {journal}
  {Physics of Atomic Nuclei}\ }\textbf {\bibinfo {volume} {79}},\ \bibinfo
  {pages} {455} (\bibinfo {year} {2016})}\BibitemShut {NoStop}%
\bibitem [{\citenamefont {Avancini}\ \emph {et~al.}(2017)\citenamefont
  {Avancini}, \citenamefont {Farias}, \citenamefont {Pinto}, \citenamefont
  {Tavares},\ and\ \citenamefont {Tim{\'o}teo}}]{avancini2017pi0}%
  \BibitemOpen
  \bibfield  {author} {\bibinfo {author} {\bibfnamefont {S.~S.}\ \bibnamefont
  {Avancini}}, \bibinfo {author} {\bibfnamefont {R.~L.}\ \bibnamefont
  {Farias}}, \bibinfo {author} {\bibfnamefont {M.~B.}\ \bibnamefont {Pinto}},
  \bibinfo {author} {\bibfnamefont {W.~R.}\ \bibnamefont {Tavares}},\ and\
  \bibinfo {author} {\bibfnamefont {V.~S.}\ \bibnamefont {Tim{\'o}teo}},\
  }\href@noop {} {\bibfield  {journal} {\bibinfo  {journal} {Physics Letters
  B}\ }\textbf {\bibinfo {volume} {767}},\ \bibinfo {pages} {247} (\bibinfo
  {year} {2017})}\BibitemShut {NoStop}%
\bibitem [{\citenamefont {Coppola}\ \emph
  {et~al.}(2020{\natexlab{b}})\citenamefont {Coppola}, \citenamefont {Dumm},
  \citenamefont {Noguera},\ and\ \citenamefont
  {Scoccola}}]{coppola2020magnetic}%
  \BibitemOpen
  \bibfield  {author} {\bibinfo {author} {\bibfnamefont {M.}~\bibnamefont
  {Coppola}}, \bibinfo {author} {\bibfnamefont {D.~G.}\ \bibnamefont {Dumm}},
  \bibinfo {author} {\bibfnamefont {S.}~\bibnamefont {Noguera}},\ and\ \bibinfo
  {author} {\bibfnamefont {N.~N.}\ \bibnamefont {Scoccola}},\ }\href@noop {}
  {\bibfield  {journal} {\bibinfo  {journal} {Journal of High Energy Physics}\
  }\textbf {\bibinfo {volume} {2020}},\ \bibinfo {pages} {1} (\bibinfo {year}
  {2020}{\natexlab{b}})}\BibitemShut {NoStop}%
\bibitem [{\citenamefont {Mei}\ \emph {et~al.}(2026)\citenamefont {Mei},
  \citenamefont {Wen}, \citenamefont {Zhou}, \citenamefont {Mao},\ and\
  \citenamefont {Huang}}]{mei2026spectral}%
  \BibitemOpen
  \bibfield  {author} {\bibinfo {author} {\bibfnamefont {J.}~\bibnamefont
  {Mei}}, \bibinfo {author} {\bibfnamefont {R.}~\bibnamefont {Wen}}, \bibinfo
  {author} {\bibfnamefont {M.}~\bibnamefont {Zhou}}, \bibinfo {author}
  {\bibfnamefont {S.}~\bibnamefont {Mao}},\ and\ \bibinfo {author}
  {\bibfnamefont {M.}~\bibnamefont {Huang}},\ }\href@noop {} {\bibfield
  {journal} {\bibinfo  {journal} {arXiv preprint arXiv:2601.22422}\ } (\bibinfo
  {year} {2026})}\BibitemShut {NoStop}%
\bibitem [{\citenamefont {Wess}\ and\ \citenamefont
  {Zumino}(1971)}]{wess1971consequences}%
  \BibitemOpen
  \bibfield  {author} {\bibinfo {author} {\bibfnamefont {J.}~\bibnamefont
  {Wess}}\ and\ \bibinfo {author} {\bibfnamefont {B.}~\bibnamefont {Zumino}},\
  }\href@noop {} {\bibfield  {journal} {\bibinfo  {journal} {Physics Letters
  B}\ }\textbf {\bibinfo {volume} {37}},\ \bibinfo {pages} {95} (\bibinfo
  {year} {1971})}\BibitemShut {NoStop}%
\bibitem [{\citenamefont {Witten}(1983)}]{witten1983global}%
  \BibitemOpen
  \bibfield  {author} {\bibinfo {author} {\bibfnamefont {E.}~\bibnamefont
  {Witten}},\ }\href@noop {} {\bibfield  {journal} {\bibinfo  {journal}
  {Nuclear Physics B}\ }\textbf {\bibinfo {volume} {223}},\ \bibinfo {pages}
  {422} (\bibinfo {year} {1983})}\BibitemShut {NoStop}%
\bibitem [{\citenamefont {Gell-Mann}\ \emph {et~al.}(1968)\citenamefont
  {Gell-Mann}, \citenamefont {Oakes},\ and\ \citenamefont
  {Renner}}]{gell1968behavior}%
  \BibitemOpen
  \bibfield  {author} {\bibinfo {author} {\bibfnamefont {M.}~\bibnamefont
  {Gell-Mann}}, \bibinfo {author} {\bibfnamefont {R.~J.}\ \bibnamefont
  {Oakes}},\ and\ \bibinfo {author} {\bibfnamefont {B.}~\bibnamefont
  {Renner}},\ }\href@noop {} {\bibfield  {journal} {\bibinfo  {journal}
  {Physical Review}\ }\textbf {\bibinfo {volume} {175}},\ \bibinfo {pages}
  {2195} (\bibinfo {year} {1968})}\BibitemShut {NoStop}%
\bibitem [{\citenamefont {Agasian}\ and\ \citenamefont
  {Shushpanov}(2001)}]{agasian2001gell}%
  \BibitemOpen
  \bibfield  {author} {\bibinfo {author} {\bibfnamefont {N.~O.}\ \bibnamefont
  {Agasian}}\ and\ \bibinfo {author} {\bibfnamefont {I.~A.}\ \bibnamefont
  {Shushpanov}},\ }\href@noop {} {\bibfield  {journal} {\bibinfo  {journal}
  {Journal of High Energy Physics}\ }\textbf {\bibinfo {volume} {2001}},\
  \bibinfo {pages} {006} (\bibinfo {year} {2001})}\BibitemShut {NoStop}%
\bibitem [{\citenamefont {Bijnens}\ and\ \citenamefont
  {Ecker}(2014)}]{bijnens2014mesonic}%
  \BibitemOpen
  \bibfield  {author} {\bibinfo {author} {\bibfnamefont {J.}~\bibnamefont
  {Bijnens}}\ and\ \bibinfo {author} {\bibfnamefont {G.}~\bibnamefont
  {Ecker}},\ }\href@noop {} {\bibfield  {journal} {\bibinfo  {journal} {Annual
  Review of Nuclear and Particle Science}\ }\textbf {\bibinfo {volume} {64}},\
  \bibinfo {pages} {149} (\bibinfo {year} {2014})}\BibitemShut {NoStop}%
\bibitem [{\citenamefont {Schwinger}(1951)}]{Schwinger:1951nm}%
  \BibitemOpen
  \bibfield  {author} {\bibinfo {author} {\bibfnamefont {J.~S.}\ \bibnamefont
  {Schwinger}},\ }\href {https://doi.org/10.1103/PhysRev.82.664} {\bibfield
  {journal} {\bibinfo  {journal} {Phys. Rev.}\ }\textbf {\bibinfo {volume}
  {82}},\ \bibinfo {pages} {664} (\bibinfo {year} {1951})}\BibitemShut
  {NoStop}%
\bibitem [{\citenamefont {Detmold}\ \emph {et~al.}(2009)\citenamefont
  {Detmold}, \citenamefont {Tiburzi},\ and\ \citenamefont
  {Walker-Loud}}]{detmold2009extracting}%
  \BibitemOpen
  \bibfield  {author} {\bibinfo {author} {\bibfnamefont {W.}~\bibnamefont
  {Detmold}}, \bibinfo {author} {\bibfnamefont {B.~C.}\ \bibnamefont
  {Tiburzi}},\ and\ \bibinfo {author} {\bibfnamefont {A.}~\bibnamefont
  {Walker-Loud}},\ }\href@noop {} {\bibfield  {journal} {\bibinfo  {journal}
  {Physical Review D---Particles, Fields, Gravitation, and Cosmology}\ }\textbf
  {\bibinfo {volume} {79}},\ \bibinfo {pages} {094505} (\bibinfo {year}
  {2009})}\BibitemShut {NoStop}%
\bibitem [{\citenamefont {Navas}\ \emph {et~al.}(2024)\citenamefont {Navas}
  \emph {et~al.}}]{ParticleDataGroup:2024cfk}%
  \BibitemOpen
  \bibfield  {author} {\bibinfo {author} {\bibfnamefont {S.}~\bibnamefont
  {Navas}} \emph {et~al.} (\bibinfo {collaboration} {Particle Data Group}),\
  }\href {https://doi.org/10.1103/PhysRevD.110.030001} {\bibfield  {journal}
  {\bibinfo  {journal} {Phys. Rev. D}\ }\textbf {\bibinfo {volume} {110}},\
  \bibinfo {pages} {030001} (\bibinfo {year} {2024})}\BibitemShut {NoStop}%
\end{thebibliography}%


\end{document}